\documentclass{article}[fulltext]
\usepackage{fullpage}
\usepackage{siunitx,algorithm,graphicx,amsmath, amssymb,url}
\usepackage[noend]{algpseudocode}
\usepackage[nocompress]{cite}
\usepackage[english]{babel}
\newcommand{\ignore}[1]{}
\usepackage{color}
\usepackage[shortlabels]{enumitem}
\usepackage{caption}
\usepackage{xr}
\newcommand{\vir}[1]{``#1"}

\begin{document}


\title{Learned Sorted Table Search and Static Indexes in Small Model Space \footnote{An extended abstract related to this paper has been presented at 20th International Conference of the Italian Association for Artificial Intelligence (AixIA 2021). }}

\author{Domenico Amato$^1$\\
	\and
	Giosu\'e Lo Bosco$^1$\footnote{corresponding author, email: giosue.lobosco@unipa.it}\\
	\and
	Raffaele Giancarlo$^1$}

\date{
	$^1$Dipartimento di Matematica e Informatica\\ 
	Universit\'a degli Studi di Palermo, ITALY\\
	\today
}

\maketitle 

\begin{abstract}

Machine Learning Techniques, properly combined with Data Structures, have resulted in  Learned Static Indexes, innovative and powerful tools that speed-up Binary Search, with the use of additional space with respect to the table being searched into. Such space is devoted to the Machine Learning Model. Although in their infancy,  they are methodologically and practically important, due to the pervasiveness of Sorted Table Search procedures. 
In modern applications, model space is a key factor and, in fact,  a major open question concerning this area is to assess to what extent one can enjoy the speed-up of Binary Search achieved by Learned Indexes while using constant or nearly constant space  models. 
In this paper, we investigate the mentioned question by (a) introducing two new models, i.e., the Learned $k$-ary Search Model and the Synoptic Recursive Model Index, respectively; (b) systematically exploring  the  time-space trade-offs of a hierarchy of existing models, i.e., the ones in the reference software platform \emph{Searching on Sorted Data},  together with the new ones proposed here. 
We document a novel  and  rather complex time-space trade-off picture, which is very informative for users as well as designers of Learned Indexing Data Structures. By adhering and extending the current benchmarking methodology, we experimentally show that the Learned $k$-ary Search Model can speed up Binary Search in constant additional space. Our second model, together with the  bi-criteria Piece-wise Geometric Model index,  can achieve a speed-up of Binary Search with a model space of $0.05\%$ more than the one taken by the table, being competitive in terms of time-space trade-off with existing proposals. The Synoptic Recursive Model Index and the bi-criteria Piece-wise Geometric Model complement each other quite well across the various levels of the internal memory hierarchy. Finally, our findings stimulate research in this area, since they highlight the need for further studies regarding the time-space relation in Learned Indexes. 

\end{abstract}

\section{Introduction}

With the aim of obtaining time and space improvements in classic Data Structures, an emerging trend is to combine
Machine Learning techniques with the ones proper of Data Structures. This new research area goes under the name of \emph{Learned Data Structures}, and it has been initiated in 2018 by Kraska et al. \cite{kraska18case}.
In particular, in such a paper the Learned Data Structures have been used mainly for the case of searching in sorted sets. This particular problem can be solved in classic algorithmics by using a well-known and optimal routine, i.e. Binary Search \cite{KnuthS, Aho1974TheDA}, or more sophisticated Data Structures, e.g. classic Indexes such as B-Trees \cite{comer1979ubiquitous}. Usually, the classic approaches consider all of the element positions in a sorted list as possible candidates to be an answer to a search query. Such an initial list is then refined in at most $O(\log n)$ iterations, where $n$ is the size of the sorted set. The main novelty in the Learned Data Structures paradigm  is the use of a Machine Learning Model trained over the elements of a sorted set, that can learn the dataset distribution. This Model uses such a knowledge  to make a prediction of the position of the query element in the sorted set. The prediction may be inaccurate, so the Model returns an interval to search into that accounts for prediction errors. As a consequence, the output of the Model is an interval of positions, where to search into. The better the Model, the smaller the interval. The final search stage on the reduced table positions interval is performed via, for the sake of exposition, Binary Search. This particular kind of Learned Data structure is referred to as Learned Index and is the main object of this research. In what follows, in order to place our contributions in the proper Literature context, we provide a brief Literature review,  followed by a road map of the paper highlighting also our contributions.

\subsection{Literature Review}

Although \emph{Learned Data Structure} is a very novel research field, it has already been extensively studied in the Literature \cite{Ferragina:2020book, Mitzenmacher22alg, Marcus20}. In what follows, we mention the main methods which can be useful for a better comprehension of the contributions provided in this paper. To this end, the most significant Learned Indexes are presented, with specific reference to their training procedures and relative benchmarking studies. Moreover, examples of real-world applications of Learned Indexes are provided, the important aspect of time/space correlation is highlighted, and for completeness, examples of other Learned Data Structures different from Learned Indexes are given.
However, The presentation is intended to be synoptic, since the interested reader can find details in the papers that are mentioned, including also a recent review on the subject \cite{Ferragina:2020book}.

\subsubsection{Core Methods and Benchmarking Platform}

The Recursive Model Index \cite{kraska18case} ({\bf RMI} for short) is the first Learned Index proposal. It is a hierarchical model which estimates the distribution of the data via a top-down approach. It can be considered as a tree-like structure, where the nodes are generic models, ranging from Neural Network Models \cite{amato2022neural} to simple linear or polynomial regression models \cite{kraska18case}. Given a query element, the internal nodes at each level identify the index of the next model (node) to use in the hierarchy. Finally, leaves provide a reduced interval to search into.
The tree structure of the {\bf RMI} is characterized by the number of levels, the number of nodes for each level and the kind of models adopted at each node. As a consequence, the {\bf RMI} depends on a consistent number of hyperparameters, whose estimation could be a serious issue in real-world contexts, as highlighted by Maltry et al. \cite{Mailtry21}.
To overcome these difficulties, Marcus et al. provide a platform, referred to as {\bf CDFShop} \cite{Marcus20b}, that can be used to generate the code of a specific {\bf RMI}, given an input dataset and specific values of its hyperparameters. In addition, given again an input dataset, the platform can provide up to ten {\bf RMI}s.

Following the seminal proposal of the {\bf RMI}, various new versions of Learned Indexes have been designed. This is the case of the Piece-wise Geometric Model Index \cite{Ferragina:2020pgm} ({\bf PGM} for short) that estimates the data distribution in a bottom-up fashion, by a Piece-wise Linear Approximation Algorithm \cite{Chen2012ApproximatingPB}. Differently from the {\bf RMI}, it is based on only one hyperparameter $\epsilon$, which represents the maximum error admitted for the index prediction. Note that despite the value of $\epsilon$ guarantees an upper bound on the search time, it does not provide any bound suggestion on the additional space used by any Learned Index with respect to the size of the input data. 

The {\bf FIT}ing-Tree Model by Kraska et al. \cite{Kraska:FITing} has been designed to overcome the mentioned space issue. It is an extension of the {\bf PGM} using the maximum number of approximation segments as an additional parameter, so that it is possible to compute the maximum additional space used by this Model. Although characterized by this new space bound property, it is not considered in this study because of its poor performance in terms of query time with respect to others Learned Indexes, as remarked in the Literature \cite{Kipf19}. 

The Radix Spline Index \cite{Kipf20} ({\bf RS} for short) is another example of a bottom-up approach to Learned Indexing, that in a different manner from the {\bf PGM} estimates the distribution through a spline curve \cite{neumann2008}. As for the {\bf FIT}ing-Tree Model, both search time and space can be controlled through two hyperparameters, i.e., the maximum error $\epsilon$ and the number of bits needed to index the spline points. However, we anticipate that such a control of space is rather poor, as documented by our experiments.

Except for the {\bf PGM}, all the  Learned Indexes mentioned so far are static and need to be rebuilt in the case the input dataset changes. Such a reconstruction could affect seriously the Learned Index performances, so a new class of indexes, referred to as Dynamic, have  been proposed. This is the case of the Adaptive Learned Index \cite{Ding20} ({\bf ALEX} for short), which provides a Dynamic Learned Index via an extension of the {\bf RMI}. 

Due to the high number of Learned Index proposals, it is evident that it is necessary to determine the strengths and weaknesses of each method. To this end, Marcus et al.\cite{Marcus20} provide an exhaustive benchmarking study of the main Learned Indexes on real datasets, supported by the development of a software platform referred to as \emph{Searching on Sorted Data} \cite{Kipf19} ({\bf SOSD} for short). The mentioned study approaches the question by considering only Binary Search as the final level of Learned Indexing. Additional pros/cons studies are available at \cite{Amato2022SPE, amato2021lncs}, considering also different types of search procedures, such as Uniform Binary Search and $k$-ary Search. However, it is evident that no clear winner emerges, across the many datasets and search routines used for experimentation. It is also evident that, as also summarized in a web platform \cite{IndexBoard}, that in most cases the best performing indexes are the {\bf RMI}, {\bf PGM} and {\bf RS}. As a consequence, these three Learned Indexes are the ones considered in this paper as a baseline to compare against.

\subsubsection{Applications}

Classic Indexes are widely used in a variety of real-world contexts, such as Databases \cite{rao1999cache} and Search Engines \cite{Morin17}. As a consequence, Learned Indexes can also make improvements in various related applications. In particular, they are widely used for Databases, providing new challenges and opportunities \cite{Wang16}, such as the development of the so-called \emph{Learned Databases} \cite{kraska2021sagedb}. They have been applied also in specific kinds of Databases, such as spatial  \cite{Li20Lisa, Wang19spatial} and biological \cite{olha2022learned} ones.
Finally, another very recent application is the development of frameworks for optimizing Database queries \cite{marcus2019neo, Zhang21, Marcus22Bao,Mikhaylov22query}.

\subsubsection{Additional Learned Data Structures}

Analogously to Learned Indexes, many methods can benefit from the combined approach of Machine Learning and classic Data Structures. An example that has been extensively discussed in the Literature is the case of the Bloom Filters \cite{Bloom1970}, whose learned version is introduced by Kraska et al. \cite{kraska18case}, extended with several variants in \cite{Mitz18, vaidya2020partitioned, dai2020adaptive} and more in-depth analysed by Fumagalli et al. \cite{Fumagalli:2021}. 
Other examples are the Learned Hash Functions  \cite{kraska18case, Singh22hash}, Learned Binary Trees  \cite{Lin22tree} and Learned Rank/Select Dictionaries \cite{Boffa21}.
However, the importance of using a Learning phase to improve the performance of a classic algorithm has not been limited only to those concerned with searching in sorted sets, but recently also for caching, scheduling, counting on data streams \cite{Mitzenmacher22alg},  and in the specific case of sorting operations \cite{kristo20sort}.

\subsubsection{An Overlooked Issue: Time/Space Correlation in Learned Indexing}

As we have mentioned, all Learned Indexes proposal offer some kind of time/space trade-off. However, this aspect has not been investigated in depth and rigorously, following the methodology coming from Classic Data Structures \cite{KnuthS}. Moreover, it is missing an assessment of how good would be constant space models at speeding-up search procedures. Indeed, two related fundamental questions have been overlooked, which are stated here: 

\begin{itemize}
    \item to what extent one can enjoy the speed-up of the search procedures provided by Learned Indexes with respect to the additional space one needs to use.
    \item how space-demanding should be a predictive model in order to speed up those procedures.
\end{itemize}

The main contribution of this paper is to provide answers to those two questions, putting this new algorithmic methodology at par with the classic one, i.e. the study of, first, the constant space models and then of the more space-demanding ones.

\subsection{Road Map of the Paper}

Here we provided a road map of the paper and our contributions.

Section \ref{sec:PS} is dedicated to a formal definition of the search on sorted data problem, also proposing via a classic solution, i.e., Binary Search. Then, we provide and discuss a very simple approach to learning from data to speed-up searching in sorted tables. Moreover, we propose a classification of Learned Indexes that includes two new ones as well as some that are leaders in the Literature, i.e., {\bf RMI}, {\bf RS} and {\bf PGM}. In particular, the first new model, referred to as \emph{Learned $k$-ary Search} ({\bf KO-US} for short), uses constant space while the other new model referred to as \emph{Synoptic RMI} ({\bf SY-RMI} for short), uses a user-defined amount of space. Apart from the novelty of the proposed models, the classification is new and methodologically important, since it allows us to systematically and coherently study whether we can obtain Learned Indexes with small space occupancy, i.e, close to constant such as a classic Binary Search, with the characteristic of being able to speed-ups Sorted Table Search procedures.

Section \ref{sec:ExpMeth} provides our experimental methodology, which extends the one recommended in the benchmarking study by Marcus et al.\cite{Marcus20}. In particular, in order to provide an evaluation of how Learned Indexes perform when the input table fits the different levels of the internal memory hierarchy, we have extended the datasets used in the benchmarking study. This is another methodologically important contribution of this scientific research.

Section \ref{sec:LM} describes and analyses the training phase of the two novel models. In particular, we focus on how the \emph{Synoptic RMI} is able to learn, in small space, key features of a variety of real datasets for the purpose of prediction. Moreover, we report useful indications, overlooked so far in the Literature, for Learned Indexes designers and practitioners about model training across different memory levels, shedding additional light on the training phase of the {\bf RS} and the {\bf PGM}. It is useful to recall that the {\bf RS} is superior to the {\bf PGM} in training time on large datasets \cite{Kipf20}. Here we show that, on small datasets, this is no longer the case.

Section \ref{sec:qc} describes and analyzes the Learned Indexes query phase, providing the main contributions of this paper. In particular, concerning the additional space, we analyse two possible cases: constant or nearly constant and parametric.

For the case of constant space, our main contribution is the study of the performance of the Learned $k$-ary Search Model in comparison with a Cubic Regression Model and the Binary Search alone. Indeed, we anticipate that the Learned $k$-ary Search Model performs better than the Binary Search alone and of the Cubic Model,  except in the case when the dataset distribution is very complex to approximate. This issue represents the main weakness of constant space models. In addition, the Learned $k$-ary Search Model has been compared with a top performing Biunary Search routine that uses a layout other than sorted, i.e., the Eytzinger Layout \cite{Morin17}. Our findings provide evidence that the Eytzinger Layout, when possible to use, is always competitive with respect to all the Models with constant or nearly constant space, even the Learned $k$-ary Search. Unfortunately, as indicated in what follows, such a lyaout cannot be used within the current Leaerned Indexing paradigm. 

For the case of parametric space, we provide a confirmation and an extension of the findings provided in the benchmarking study by Marcus et al.\cite{Marcus20}. Indeed, the  new model introduced in the study, i.e. the Synoptic {\bf RMI},  and the bi-criteria {\bf PGM}, perform better than the Binary Search alone, across all the datasets and memory levels, using very small additional space with respect to the input table. Moreover, even the most complex models, excluding the {\bf RS} on the lower memory levels, achieves very good performance considering a bound of at most 10\% of additional space. It is also competitive with respect to the bi-criteria {\bf PGM}. 
We investigate also Parametric Models' time and space relationships, showing that while their query times can differ by constant factors, the corresponding spaces can disagree by several orders of magnitude. The main finding is that space seems to be the real key to the Model's efficiency. 
This provides additional insights into the time/space relationship of Learned Indexes, with respect to what is known in the Literature. 

Finally, our analysis also provides useful guidelines  to the practitioners interested in using Learned Indexes. 

\begin{figure}[tbh]
	\centering
	\includegraphics[scale=0.8]{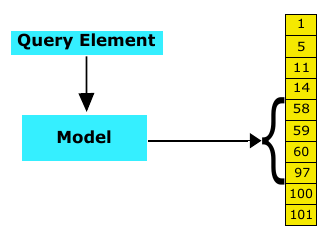}
	\caption{{\bf  A general paradigm of  Learned Searching in a Sorted Table \cite{Marcus20}}. The model is trained on the data in the table. Then, given a query element, it is used  to predict the interval in the table where to search (included in brackets in the figure).}
	\label{fig:Par}
\end{figure}

\section{Learning from a Static Sorted Set to Speed-Up Searching in It}\label{sec:PS}
Consider a sorted table $A$ of $n$ keys, taken from a universe $U$.  It is well known that Sorted Table Search can be phrased as the Predecessor Search Problem:  for a given query element $x$, return the $A[j]$ such that $A[j] \leq x < A[j+1]$. With reference to such a problem, in the following, we describe the classic solutions in the Literature and how to transform it into a learning-prediction one.

\subsection{Solution with a Sorted Search Routine}\label{sec:methods}

It is well-known in Algorithmics \cite{KnuthS, Cormer2009, Peterson57, Aho1974TheDA} that the Predecessor Search Problem can be solved with Sorted Table Search routines, such as Binary and Interpolation Search. For the aim of this paper and according to the benchmarking study, we use the C++ {\bf lower\_bound} routine, denoted as {\bf BS} and informally referred to as Standard. In addition to this method, we use the best routines that come out of the study by Khuong and Morin \cite{Morin17}, i.e., Uniform Binary Search \cite{KnuthS}, denoted as {\bf US}, and the Eytzinger Layout Search, denoted as {\bf EB}. For the convenience of the reader, details about all the above-mentioned search procedures are in Section \ref{S-sec:BS} of the Supplementary File. We anticipate that other routines may be considered in this study, such as Interpolation Search or its variant TIP \cite{VanSandt19}, but the extensive experiments conducted in \cite{amatoPhd22} show that they are not competitive in the Learned Indexing scenario. Therefore, in order to keep this paper focused on relevant contributions, they are omitted here.

\subsection{Learning from Data to Speed-Up Sorted Table Search: A simple View with an Example}\label{sec:learn}

Kraska et al. \cite{kraska18case} have proposed an approach that transforms the Predecessor Search problem into a learning-prediction one. With reference to Figure \ref{fig:Par}, the model learned from the data is used as a predictor of where a query element may be in the table. To fix ideas, Binary Search is then performed only on the interval returned by the model. 

\begin{figure}[tbh]
	\centering
	(a)
	\begin{minipage}{0.25\textwidth}
		\includegraphics[width=\linewidth]{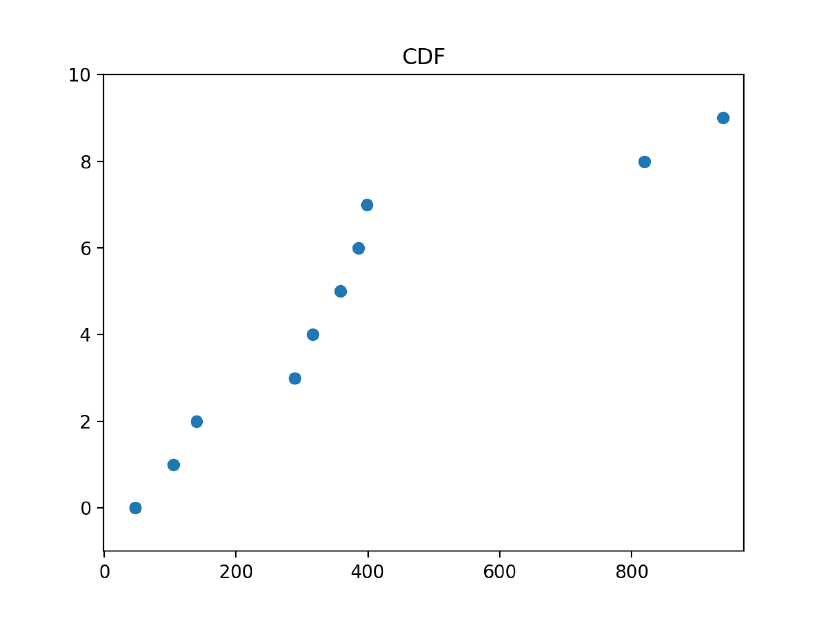}
	\end{minipage}\hfill
	(b)
	\begin{minipage}{0.25\textwidth}
		\includegraphics[width=\linewidth]{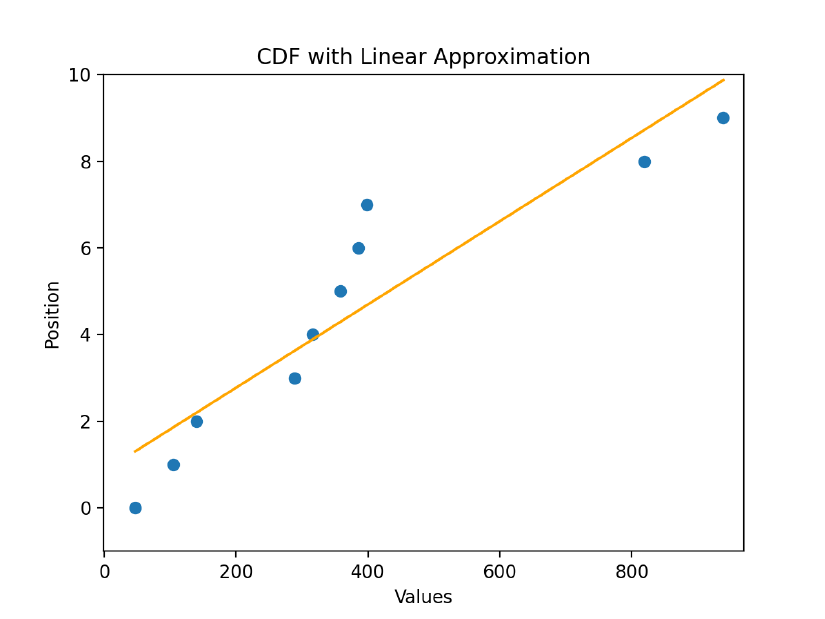}
	\end{minipage}\hfill
	(c)
	\begin{minipage}{0.25\textwidth}%
		\includegraphics[width=\linewidth]{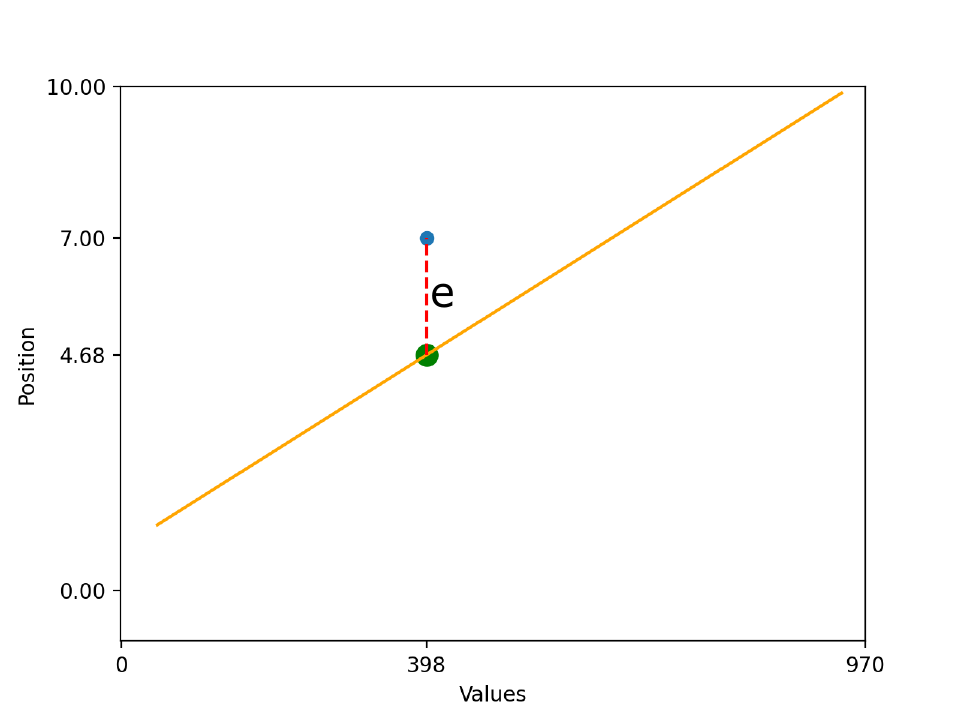}
	\end{minipage}
	\caption{{\bf The Process of Learning a Simple Model via Linear Regression.} Let the table A be $[47, 105, 140, 289, 316, 358, 386, 398, 819, 939]$. (a) the empirical \emph{CDF} of $A$;  (b) the line (in orange) associated with a linear model obtained via Linear Regression; and (c) the error $e$ made by the model in predicting the query element $398$.}

	\label{fig:CDF}
\end{figure}

We now outline the simplest technique that can be used to build a model for $A$, providing also an example. It relies on Linear Regression,  with Mean Square Error Minimization \cite{FreedmanStat}. We start with the example. Consider Figure \ref{fig:CDF} and the table $A$ in the caption.

\begin{itemize}
    \item {\bf Ingredient One of Learned Indexing: The Cumulative Distribution Function (\emph{CDF} for Short) of a Sorted Table}. With reference to Figure \ref{fig:CDF}(a), we can plot the elements of $A$ in a graph, where the abscissa reports the value of the elements in the table and the ordinates are their corresponding ranks. The result of the plot is reminiscent of a discrete Cumulative Distribution Function \emph{CDF} that underlines the table. The specific  construction exemplified here can be generalized to any sorted table, as discussed in Marcus et al.\cite{Marcus20}. In the Literature, for a given table, such a discrete  curve is referenced as \emph{CDF}.
    \item {\bf Ingredient Two of Learned Indexing: A Model for the \emph{CDF}}. Now, it is essential to transform the discrete \emph{CDF} into a continuous curve. The simplest way to do this is to fit a straight line of equation $F(x)=ax+b$ to the \emph{CDF} (this process is shown in Figure \ref{fig:CDF}(b)). In this example, we use Linear Regression with Mean Square Error Minimization in order to obtain $a$ and $b$. They are 0.01 and 0.85, respectively. 
    
    \item {\bf Ingredient Three of Learned Indexing: The Model Error Correction}. Since $F$ is an approximation of the ranks of the elements in the table, applying it to an element in order to predict its rank, we may produce an error $e$. With reference to Figure \ref{fig:CDF}(c), applying the model to the element $398$, we obtain a predicted rank of $4.68$, instead of $7$, which is the real rank. So, the error made by the model $F(x)=0.01*x+0.85$ on this element is $e = 7 - \lceil 4.68 \rceil = 2$. Therefore, in order to use the equation $F$ to predict where an element $x$ is in the table, we must correct for this error. Indeed, we consider the maximum error $\epsilon$ computed as the maximum distance between the real rank of the elements in the table and the corresponding rank predicted by the model. The maximum error $\epsilon$ is used to set the search interval of an element $x$ to be $[F(x) - \epsilon, F(x) + \epsilon]$. In the example we are discussing, $\epsilon$ is 3.
\end{itemize}

More in general, in order to perform a query, the model is consulted and an interval in which to search is returned. Then,  Binary Search on that interval is performed. Different models may use different schemes to determine the required range, as outlined in Section  \ref{sec:models}. 
The reader interested in a rigorous presentation of those ideas can consult Marcus et al. \cite{Marcus20b}. 
In this paper, we characterize the accuracy in the prediction of a model via the \emph{reduction factor}: the percentage of the table that is no longer considered for searching after the prediction of a rank. Because of the diversity across models to determine the search interval, and in order to place all models on a par, we estimate empirically the reduction factor of a model. That is, with the use of the model and over a batch of queries, we determine the length of the interval to search into for each query. Based on it, it is immediate to compute the reduction factor for that query. Then, we take the average of those reduction factors over the entire set of queries as the reduction factor of the model for the given table.

\subsection{A Classification of Learned Indexing Models}\label{sec:models}

\begin{figure}[!htb]
    \begin{minipage}{0.15\textwidth}
	\end{minipage}\hfill
	(a)
	\begin{minipage}{0.3\textwidth}
		\includegraphics[width=\linewidth]{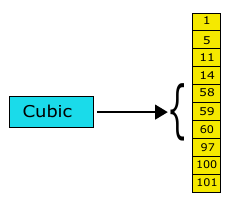}
	\end{minipage}\hfill
	(b)
	\begin{minipage}{0.3\textwidth}%
		\includegraphics[width=\linewidth]{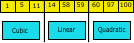}
	\end{minipage}\hfill
	 \begin{minipage}{0.15\textwidth}
	\end{minipage}\hfill
	\caption{ {\bf  Examples of Various Learned Indexes that Use Constant Space}. (a) an Atomic Model, where  the box Cubic means that the CDF of the entire dataset is estimated by a cubic function via Regression, in analogy with the linear approximation exemplified  in Figure \ref{fig:CDF}. (b) An example of a {\bf KO-US}, with $k=3$. The top part divides the table into three segments and it is used to determine the model to pick at the second stage. Each box indicates which Atomic  Model is used for prediction on the relevant portion of the table.}  
	\label{fig:constmodels}
\end{figure}

\begin{figure}[!htb]
	(a)
	\begin{minipage}{0.29\textwidth}
		\includegraphics[width=0.9\linewidth]{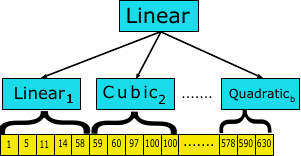}
	\end{minipage}\hfill
	(b)
	\begin{minipage}{0.29\textwidth}%
		\includegraphics[width=0.6\linewidth]{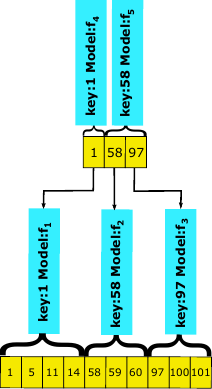}
	\end{minipage}\hfill
	(c)
	\begin{minipage}{0.29\textwidth}%
		\includegraphics[width=\linewidth]{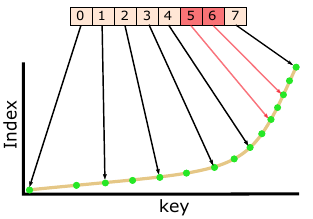}
	\end{minipage}\hfill
	\caption{ {\bf  Examples of Various Learned Indexes that Use Space in Fuction of Some Parameters} (see also \cite{Marcus20}). (a)  An example of an {\bf RMI} with two layers and branching factor equal to $b$. The top box indicates that the lower models are selected via a linear function. As for the leaf boxes,  each  indicates which Atomic  Model is used for prediction on the relevant portion of the table.   (b)  An example of  a {\bf PGM }  Index. At the bottom, the table is divided into three parts. A new table is so constructed and the process is iterated. (c) An example of  an {\bf RS } Index. At the top, the buckets where elements fall, based on their three most significant digits. At the bottom,  a linear spline approximating the CDF of the data, with  suitably chosen spline points. Each bucket points to a spline point so that, if  a query element falls in a bucket  (say six), the search interval is limited by the spline points pointed to by that bucket and the one preceding it (five in our case).}  
	\label{fig:parammodels}
\end{figure}

With the exception of the Eytzinger Binary Search, all procedures mentioned in Section \ref{sec:methods} have a natural Learned version.  Indeed, all models currently known in the Literature naturally fit sorted table layouts for the final search stage  but, for that purpose,  array layouts other than sorted or more complex Data Structures cannot be used. Given a Learned version of the two mentioned procedures, its time and space performances depend critically on the model used to predict the interval to search into. Here we propose a classification of Models that comprises four classes. The first two, shown in Figure \ref{fig:constmodels}, consist of models that use constant space, while the other two, shown in Figure \ref{fig:parammodels}, consist of models that use space as a function  of some model parameters.  
For each of them, the reduction factor is determined as described in Section \ref{sec:PS}. Moreover, as already pointed out, the Learned $k$-ary Search   and the Synoptic {\bf RMI} Models are new  and  fit quite naturally in the classification that we present. 

\paragraph{Atomic Models: One Level and no Branching Factor}\label{sec:two-level}

\begin{itemize}
	\item {\bf Simple Regression}\cite{FreedmanStat}.  We use linear, quadratic and  cubic regression models. Each can be thought of as an atomic model in the sense that it cannot be divided into \vir{sub-models}. Figure \ref{fig:constmodels}(a) provides  an example.  We report that the most appropriate regression model in terms of query times and reduction factor is the cubic one. We omit those results for brevity and to keep our contribution focused on the important findings. However, they can be found in \cite{amatoPhd22}. For this reason, the cubic model, indicated in the rest of the manuscript by {\bf C},  is the only one that is included in what follows.
\end{itemize}

\paragraph{A Two-Level Hybrid Model, with Constant Branching Factor}

\begin{itemize}
    
    	\item {\bf KO-US: Learned  k-ary Search}.  This model partitions the table into a fixed number of segments, bounded by a small constant, i.e. at most 20 in this study, in analogy with a single iteration of the  k-ary Search routine \cite{Schlegel09, Shutz18}. An example is provided in Figure \ref{fig:constmodels}(b). For each segment,  Atomic Models are computed  to approximate the \emph{CDF} of the table elements in that segment. Finally, the model that guarantees the best reduction factor is assigned to each segment. As for the prediction, a sequential search is performed for the second level segment to pick and the corresponding model is used for the prediction, followed by  Uniform Binary Search, since it is superior to the Standard one (data not reported and available upon request). We anticipate that for the experiments conducted in this study, $k$ has been chosen in the interval $[3,20]$. For conciseness, only results for the model with $k=15$ are reported, since it is the value with the best performance in terms of query time (data not reported and available upon request). Accordingly, from now on, {\bf KO-US} indicates the Model with $k = 15$.
\end{itemize}

\paragraph{Two-Level RMIs with Parametric Branching Factor}\label{sec:parrmi}

\begin{itemize}
	\item {\bf Heuristically Optimized RMIs.}  Informally, an {\bf RMI}  is a multi-level, directed graph, with Atomic Models at its nodes. When searching for a given key and starting with the first level, a prediction at each level  identifies the model of the next level to use for the next prediction. This process continues until a final level  model is reached. This latter is used to predict the table interval to search into. As pointed out in the benchmarking study,  in most applications, a generic {\bf RMI} with  two layers, a tree-like  structure and a branching factor $b$ suffices. An  example is  provided in Figure \ref{fig:parammodels}(a). 
	It is to be noted that Atomic Models are {\bf RMI}s. Moreover, the difference between Learned k-ary Search and {\bf RMI}s is that the first level in the former partitions the table, while that same level in the latter partitions the Universe of  the elements. Following the benchmarking study and for a given table, we use two-layers {\bf RMI}s that we obtain using the optimization software provided in {\bf CDFShop}, that returns up to ten versions of the generic {\bf RMI}, for a given input table. That is, for each model, the optimization software picks an appropriate branching  factor and the type of regression to use within each part of the model, those latter quantities being the parameters that control the precision of its prediction as well as its space occupancy. It is also to be remarked, as pointed out in \cite{Marcus20b},  that the optimization process provides only approximations to the real optimum and it is heuristic in nature, with  no theoretic approximation performance guarantees. The problem of finding an optimal model in polynomial time is open.  
	
	\item {\bf SY-RMI: A Synoptic RMI.} For a given set of tables of approximately the same size, we use {\bf CDFShop} as above to obtain a set of models (at most 10 for each table). For the entire set of models so obtained and each model in it, we compute the ratio (branching factor)/(model space) and we take the median of those ratios as a measure of branching factor {\em per unit} of model space, denoted $UB$. Among the {\bf RMI}s returned by {\bf CDFShop}, we pick the relative majority winner, i.e., the one that provides the best query time, averaged  over a set of simulations. When one uses such a model on tables of approximately the same size as the ones used as input to {\bf CDFShop}, we set the branching factor to be a multiple of $UB$, that depends on how much space the model is expected to use relative to the input table size.  Since this model can be intuitively considered as the one that best summarizes the output of {\bf CDFShop} in terms of query time, for the given set of tables.  
 The final model is informally referred to as Synoptic.  
\end{itemize}

\paragraph{CDF Approximation-Controlled Models}

\begin{itemize}
	\item  {\bf PGM} \cite{Ferragina:2020pgm}. It is also a multi-stage model, built bottom-up and queried top down.  It uses a user-defined approximation parameter $\epsilon$, that controls the prediction error at each stage. With reference to Figure \ref{fig:parammodels}(b), the table is subdivided into three pieces. A prediction in each piece can be provided  via a linear model guaranteeing an error of $\epsilon$.  A new table is formed by selecting the minimum values in each of the three pieces. This new table is possibly again partitioned into pieces, in which a linear model can make a prediction within the given error. The process is iterated until only one linear model suffices, as in the case in the figure. A query is processed via a series of predictions, starting at the root of the tree.  Also in this case, for a given table, at most ten models have been built as prescribed in the benchmarking study with the use of the parameters, software and methods provided there, i.e,  {\bf SOSD}. It is to be noted that the { \bf PGM} index, in its bi-criteria version, is able to return the best query time index, within a given amount of space the model is supposed to use. Experiments are performed also with this version of the {\bf PGM}, denoted for brevity as {\bf B-PGM}. The interested  reader can find a discussion regarding more variants of this {\bf PGM} version in \cite{amatoPhd22}.

	\item {\bf RS} \cite{Kipf20}. It is a two-stage model. It also uses a user-defined approximation parameter $\epsilon$. With reference to Figure \ref{fig:parammodels}(c), a spline curve approximating the CDF of the data is built. Then, the radix table is used to identify spline points to use to refine the search interval. Also in this case, we have performed the training as described in the benchmarking study. 
	
\end{itemize}

In what follows, for ease of reference, models in the first two classes are referred to as constant space models, while the ones in the remaining classes as parametric space models.

\section{Experimental Methodology}\label{sec:ExpMeth}


Our experimental set-up follows closely the one outlined in the already mentioned benchmarking study by Marcus et al \cite{Marcus20}. Since an intent of this study is to gain deeper insights regarding the circumstances in which Learned versions of Sorted Table Search procedure and Indexes are profitable in small additional space with respect to the one taken by the input table, accross  the main memory hierarchy, we derive our own  benchmark datasets from the ones in the study by Marcus et al \cite{Marcus20}.

\subsection{Hardware}\label{hardware}
All the experiments have been performed on a workstation equipped with an Intel Core i7-8700 3.2GHz CPU with three levels of cache memory:
(a) 64kb of {\bf L1} cache; (b) 256kb of {\bf L2} cache; (c)12Mb of shared {\bf L3} cache. 
The $cls$ and $size$, defined in Section \ref{sec:parrmi} are respectively 64 and 8 bytes. The total amount of system memory is 32 Gbyte of DDR4. The operating system is Ubuntu LTS 20.04.

\subsection{Datasets}\label{sec:Datasets}


The same real datasets of the benchmarking study are used. In particular, attention is restricted to integers only, each represented with 64 bits unless otherwise specified. For the convenience of the reader, a list of those datasets, with an outline of their content, is provided next. 

\begin{itemize}
	\item {\bf amzn}: book popularity data from Amazon. Each key represents the popularity of a particular book. Although two versions of this dataset, i.e., 32-bit and 64-bit, are used in the benchmarking, no particular differences are observed in the results of our experiments, and for this reason we report only those for the 64-bit dataset. The interested reader can find the results for the 32 bits version in \cite{amatoPhd22}.
	\item {\bf face}: randomly sampled Facebook user IDs. Each key uniquely identifies a user.
	\item {\bf osm}: cell IDs from Open Street Map. Each key represents an embedded location.
	\item{\bf	wiki}: timestamps of edits from Wikipedia. Each key represents the time an edit was committed.
\end{itemize}


\begin{figure}[tbh]
\centering
		(a)
		\begin{minipage}{0.45\textwidth}
			\includegraphics[width=\linewidth]{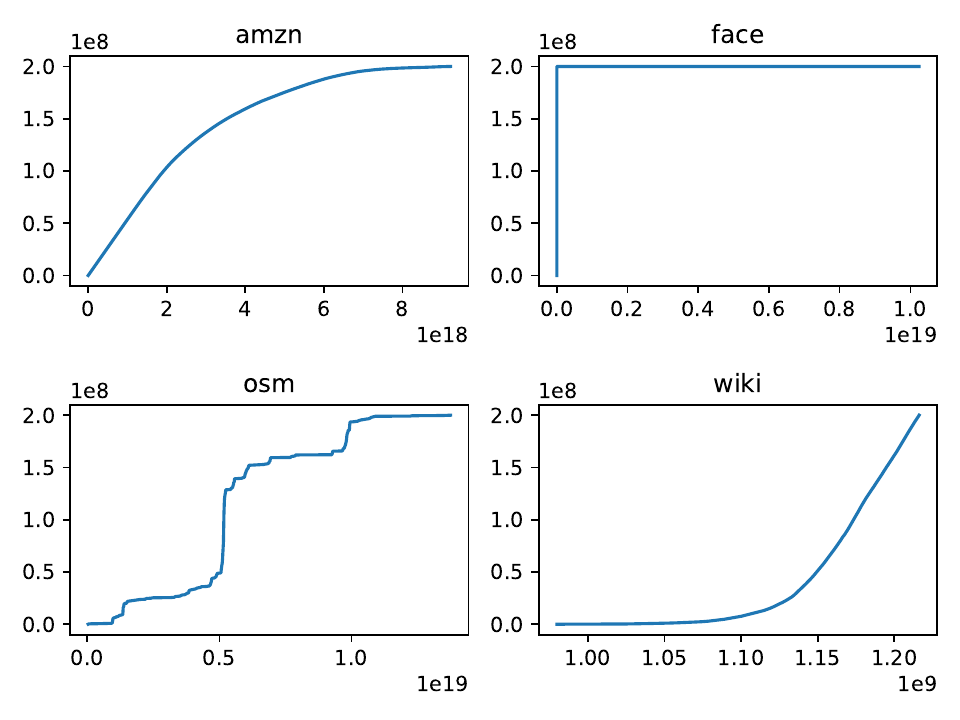}
		\end{minipage}\hfill
		(b)
		\begin{minipage}{0.45\textwidth}
			\includegraphics[width=\linewidth]{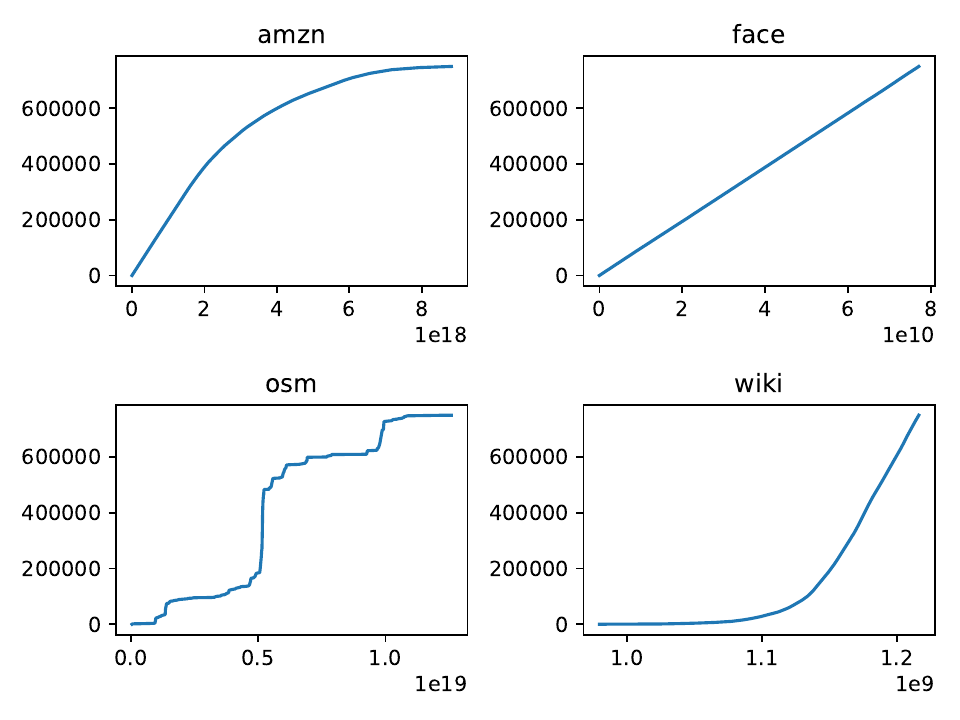}
		\end{minipage}\hfill
	\caption{{\bf  The \emph{CDF} of the Main Datasets}. For each dataset coming from the benchmarking study, the value of each of its elements is reported on the abscissa and its  position on the ordinate. In particular, Figure (a) is referred to the {\bf L4} memory level, while (b) to {\bf L3}.}
	\label{fig:dataCDF}
\end{figure}

Moreover, for the purpose of this research, as already mentioned above, additional datasets are extract from the ones just mentioned. For each of those datasets, three new ones are obtained in order to fit each lower level of the internal memory hierarchy. In particular, each new dataset is obtained by sampling the original one so that the \emph{CDF}  is similar to the original one. The interested reader can find more details of this extraction procedure in \cite{amatoPhd22}.
Letting $n$ be the number of elements in a table, for the computer architecture that is been used, the details of the generated tables are the following. 
	
	\begin{enumerate}
		\item [$\bullet$] {\bf Fitting in L1 cache: cache size 64Kb.}  Therefore, $n=3.7K $ is chosen.
		\item [$\bullet$] {\bf Fitting in L2 cache: cache size 256Kb.} Therefore, $n= 31.5K$ is chosen. 
		\item [$\bullet$] {\bf Fitting in L3 cache: cache size 8Mb.} Therefore, $n=750K$ is chosen. 
		\item [$\bullet$] {\bf Fitting in PC Main Memory (L4): memory size 32Gb.} Therefore, $n=200M$ is chosen, i.e., the entire dataset.

	\end{enumerate}

 The rationale for the choice of those datasets, in particular the ones coming from the benchmarking study, is that they provide different Empirical \emph{CDF}, as shown in Figure \ref{fig:dataCDF}(a), and this allows to measure the performance of Learned Indexes considering different possible characteristics of real-world data. It is to be noted that the {\bf face} dataset is somewhat special. Indeed, the shape of its \emph{CDF} (see Figure \ref{fig:dataCDF}(a)) is determined by 21 outliers at the end of the table:  all the elements of that dataset, up to the first outlier, have essentially the same distance between consecutive elements. That is, they are all on a straight line. This regularity breaks with the first outlier that, together with the other ones, do not follow such a nice pattern. For lower memory levels, the \emph{CDF}  of the corresponding {\bf face} datasets becomes a straight line, as exemplified in Figure \ref{fig:dataCDF}(b) for the {\bf L3} memory level. As for the remaining datasets, their smaller versions follow closely the \emph{CDF} of the biggest datasets, as again exemplified in Figure \ref{fig:dataCDF}(b) for the {\bf L3} memory level.
	
As for query dataset generation, for each of the tables built as described above, we extract uniformly and at random (with replacement) from the Universe $U$ a total of two million elements, 50\% of which are present and 50\% absent, in each table.

\section{Training of the Novel Models: Analysis and Insights into Model Training}\label{sec:LM}

We now focus on the training phase of the novel Models and we compare their performance  with the Literature standards included in this research.  In order to  assess how well a Learned Index Model can be trained, three indicators are important: the time required for learning, the reduction factor that one obtains and the time needed to perform the prediction. A quantification of the first parameter is provided and discussed here. The other two indicators are strongly dependent on each other, with the reduction factor being related to space. In turn, those two indicators affect query time. Therefore, they are best discussed in Section \ref{sec:qc}. We anticipate that our analysis of the training time performed here provides useful and novel insights into model training for Learned Indexing.  All the training experiments have been performed on the datasets mentioned in Section \ref{sec:Datasets}, across all internal memory levels.

\subsection{Mining SOSD Output for the Synoptic RMI}

\begin{figure}[t]
	\centering
	\includegraphics[width=1\textwidth]{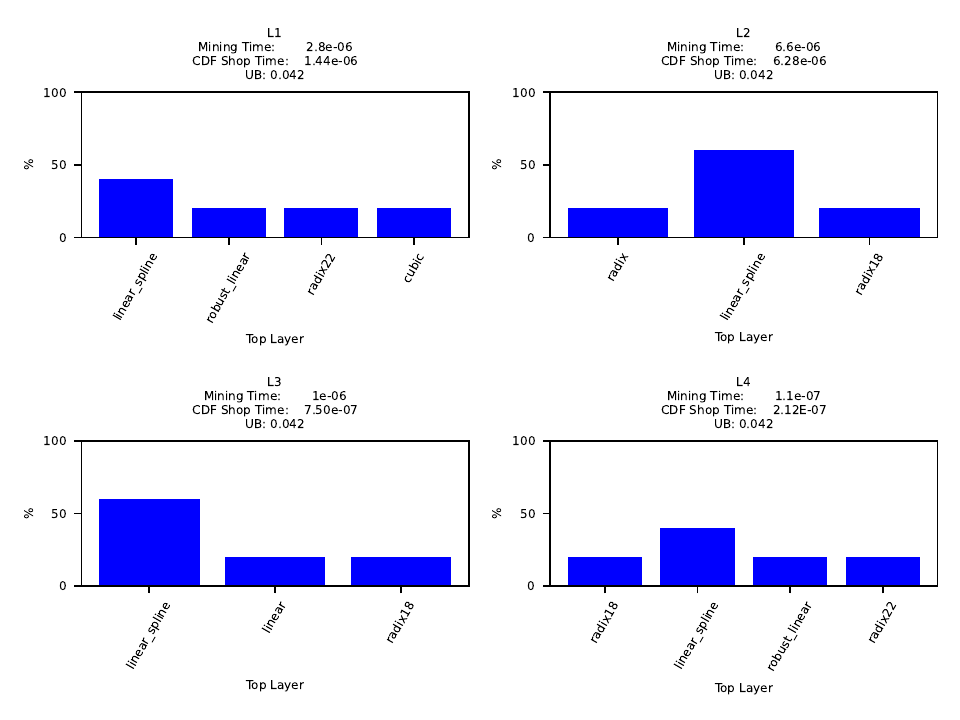}
	\caption{ {\bf Time and $UB$ for the Identification of the Synoptic RMIs.} For each memory level, only the top layer of the various models is indicated in the abscissa, while the ordinate indicates the number of times, in percentage, the given model is the best in terms of query performance on a table. 
 On top of each histogram, it is reported the branching factor per unit of space as well as the mining time to build the synoptic Models. For comparison, we also report the same time spent in obtaining the output of {\bf CDFShop.}}\label{F:TopLayer}
	
\end{figure}

As anticipated in Section \ref{sec:models}, in order to set the levels and $UB$ of the Synoptic {\bf RMI}, it is necessary to process the output of {\bf SOSD} for each dataset and memory level. Indeed, as described in Section \ref{sec:models}, once it is set a space budget for the model, the corresponding branching factor is computed by multiplying it by $UB$. In particular, we compute three versions of a Synoptic {\bf RMI} using a percentage of space of 0.05\%, 0.7\% and 2\% with respect to the input table size. With regard to the layers choice, the simulation to identify the relative majority {\bf RMI}s is performed on query datasets extracted as described in the previous Section, but using only $1\%$ of the number of query elements specified there.
The statistics regarding the results of such a simulation are summarized in Figure \ref{F:TopLayer}. In particular, for each memory level, we report the computed $UB$. Furthermore, limited to the top layer of an {\bf RMI}, we also report the models associated with the best ones. The time it took to identify the proper Synoptic {\bf RMI} (average time per element, over all {\bf RMI}s returned by {\bf CDFShop}, denoted as mining time)  is also reported, together with  the same time required to obtain  the output of {\bf CDFShop}. As it is evident, the mining time is comparable with the performance of {\bf CDFShop}.  It is also evident from that Figure that the variety of best-performing models represents well various challenges for the learning of the \emph{CDF} of real datasets. Therefore, given such a variety, it is far from obvious that the median $UB$ is the same for each memory level. Moreover, the relative majority model is also the same across memory levels, i.e., linear spline, with linear models for each segment of the second layer.

\subsection{Training Time Comparison Between Novel Models and the State of the Art}

In what follows, we divide the training time comparison into two groups: Constant and Parametric Space Models. As for the first group, we consider the new Model and only the Cubic Atomic Model,  excluding the Linear and Quadratic  ones for the reasons mentioned earlier in this paper. For the Cubic Model, the training time on a given dataset is due to the computation of its parameters via a Polynomial Regression. As for the Learned  $k$-ary Search Model, its training consists of partitioning  the table into $k$ segments. Then, for each segment, Atomic Models are used to approximate the local \emph{CDF} of the elements belonging to that  segment, and among them, the Model  with the best reduction factor is chosen.  For each dataset and each memory level, the resulting training  times  are reported in Table \ref{T:CM-TTL4} and Tables \ref{S-T:CMTTL1}-\ref{S-T:CMTTL3} of the Supplementary File. As expected, the Learned $k$-ary Search Model is slower than the Cubic Atomic Model, but the important fact is that the slowdown is due to constant multiplicative factors rather than order of magnitude. That is, the slow-down is tolerable. Another additional  and counter intuitive finding is  that the training time of both Models, on average, is better for the cases of big datasets with respect to smaller ones. We analyzed the training code in order to get insights into such a fact.  It turns out that the cost of the matrix products involved in the training computation of both models depends on the size of the involved operands. As the size of the dataset grows, such a cost is amortized on a larger and larger number of elements.

Regarding the second group, we consider the new model and the ones described in Section \ref{sec:models}, i.e., {\bf RMI}, {\bf PGM} and {\bf RS}.  The training time is computed using two different platforms: {\bf CDFShop} in the case of the Synoptic {\bf RMI} and {\bf RMI}, and {\bf SOSD} for {\bf PGM} and {\bf RS}. It is useful to recall that, in the case of the State of the Art Models, the result of a single execution of those two platforms returns a batch of up to ten models, so the reported times refer to the execution of the entire learning suite. That is, the training of those models consists of a batch of model instances from which a user can choose.  On the other hand, for the Synoptic {\bf RMI}, it is referred to the training  of a single {\bf RMI} with a given branching factor and layers composition. For each dataset and each memory level, the results are reported in Table \ref{T:PM-TTL4} and Tables \ref{S-T:PMTTL1}-\ref{S-T:PMTTL3} of the Supplementary File. The time needed to train the Synoptic 
{\bf RMI} is comparable to the one needed to train a batch of {\bf RMI} Models. This latter, as already known, being worse than the time to train a batch of {\bf RS} or {\bf PGM} Models. Such a State of the Art is not considered  problematic for the deployment of the  {\bf RMI} in application contexts and the training time of the Synoptic {\bf RMI} is in-line with the mentioned Literature Standards.  For completeness, we mention that the reason  for which  the time needed to train a unique Synoptic {\bf RMI} Model is very close to the training of a batch on {\bf RMI} Models is due to  a library start-up overhead time. Such a time is   mitigated for the case of the training of a batch of models, while it becomes dominant in training a single model. Fortunately, the {\bf CDFShop} or the {\bf SOSD} training executions are a \vir{one time only} process, in which the output can then be reused over and over again, suggesting that this overhead time is of little relevance for the case of a production environment.

\subsection{Insights into the Training Time of the {\bf RS} and {\bf PGM} Models}

Another important contribution that this research provides is a more refined assessment of the relation between the {\bf RS} and {\bf PGM} indexes, in terms of training time.  In Table \ref{T:RSPGM-TT}, for each dataset and memory level, we report the training time of those two indexes. As well discussed in \cite{Kipf20}, those two Learned Indexes can both be built in one pass over the input, with important implications, one being that they can be trained faster than the {\bf RMI}s, even with  one order of magnitude speed-up.  However, in that study  as well as in the benchmarking one, the {\bf RS} is reported as superior to the {\bf PGM} in terms of training time. It is to be noted that the datasets that they used are the largest ones in this study. With reference to  Table \ref{T:RSPGM-TT}, our experiments confirm such a finding. On the other hand, the {\bf PGM} is \vir{more effective} in terms of training time across the lower memory hierarchy.  The reason may be the following. Those two indexes both use streaming procedures in order to approximate the \emph{CDF} of the input dataset within a parameter $\epsilon$,  via the use of straight line segments that partition the Universe. The main difference between the two is that the latter finds an optimal partition, determined via a well-known algorithm (see references in  \cite{Ferragina:2020pgm}),  while the former finds a partition that approximates the optimal, as described in \cite{neumann2008}. Such an approximation algorithm is supposed to be faster than the optimal one but apparently this speed pays off on large datasets.

\begin{table}
	\begin{center}
		\caption{ {\bf Constant Space Models Training Time for L4 Tables}. The first column indicated the datasets. The remaining columns indicate the model used for the learning phase. Each entry reports the training time in seconds and per element.}\label{T:CM-TTL4}
		\small
		\begin{tabular}{|c|c|c|}
		    \hline
			 &  {\bf KO-US} & {\bf C} \\
			\hline
			
			{\bf amzn} &3.7e-08&1.4e-08\\
			\hline
			{\bf face} &3.6e-08&1.4e-08\\
			\hline
			{\bf osm} &3.6e-08&1.4e-08\\
			\hline
			{\bf wiki}  &3.6e-08&1.4e-08\\
			\hline
		\end{tabular}
	\end{center}
\end{table}

\begin{table}
	\begin{center}
		\caption{ {\bf Paramentric Models Training Time for L4 Tables}. The first column indicated the datasets. The remaining columns indicate the model used for the learning phase. In particular, each entry reports the time to train the Synoptic {\bf RMI} and an entire batch of models via the {\bf CDFShop} and  the {\bf SOSD} libraries as specified in the Main text. The time is in seconds and per element.}\label{T:PM-TTL4}
		\small
		\begin{tabular}{|c|c|c|c|c|}
		    \hline
			 & {\bf CDFShop SY-RMI 2\%} &  {\bf CDFShop RMI}& {\bf SOSD RS} & {\bf SOSD PGM}\\
			\hline
			
			{\bf amzn} &1.1e-06 &2.2e-06 &2.1e-07 &5.0e-07\\
			\hline
			{\bf face} &1.3e-06&2.5e-06 &2.1e-07 &6.5e-07\\
			\hline
			{\bf osm} &1.2e-06&2.5e-06 &2.2e-07 &7.4e-07\\
			\hline
			{\bf wiki} &1.1e-06&2.2e-06 &1.9e-07 &4.1e-07\\
			\hline
		\end{tabular}
	\end{center}
\end{table}

\begin{table}
	\begin{center}
		\caption{ {\bf Comparison between RS and PGM Training Time}. For each dataset and memory level, we report the training time for the {\bf RS} and {\bf PGM} models in seconds}\label{T:RSPGM-TT}
		\small
		\begin{tabular}{|c|c|c|c|c|c|c|c|c|}
		    \hline
		    & \multicolumn{2}{c|}{L1} & \multicolumn{2}{c|}{L2} & \multicolumn{2}{c|}{L3} & \multicolumn{2}{c|}{L4} \\ \hline
			 & {\bf SOSD RS} & {\bf SOSD PGM} & {\bf SOSD RS} & {\bf SOSD PGM} & {\bf SOSD RS} & {\bf SOSD PGM} & {\bf SOSD RS} & {\bf SOSD PGM}\\
			\hline
                {\bf amzn} &3.5e-06 &5.0e-07 &3.5e-07 &5.0e-08 &2.4e-08 &3.4e-08 &2.1e-07 &5.0e-07 \\ \hline
                {\bf face} &1.1e-06 &3.9e-07 &1.1e-07 &3.9e-08 &1.4e-08 &2.4e-08 &2.1e-07 &6.5e-07 \\ \hline
                {\bf osm} &6.9e-06 &4.0e-07 &6.9e-07 &4.0e-08 &3.5e-08 &3.8e-08 &2.2e-07 &7.4e-07 \\ \hline
                {\bf wiki} &1.0e-05 &3.7e-07 &1.0e-06 &3.7e-08 &5.1e-08 &3.7e-08 &1.9e-07 &4.1e-07 \\ \hline

		\end{tabular}
	\end{center}
\end{table}

\section{Query Experiments}\label{sec:qc}

The query experiments are performed using all the methods described in Sections \ref{sec:methods} and \ref{sec:models}. The query datasets have been generated as described in Section \ref{sec:Datasets} and the models have been trained as described in Section \ref{sec:LM}. Following that Section, we divide the  presentation of the query experiments and the relative discussion into two groups. For both groups, for conciseness, we report here only the experiments for  the {\bf amzn} and the {\bf osm} datasets since they are representative of two different levels of difficulty in learning their \emph{CDF}s. The results regarding the other datasets are reported in the Supplementary File.

\subsection{Constant Space Models}\label{sec:qc-cm}

\begin{figure}[t]
	\centering
	\includegraphics[width=1\textwidth]{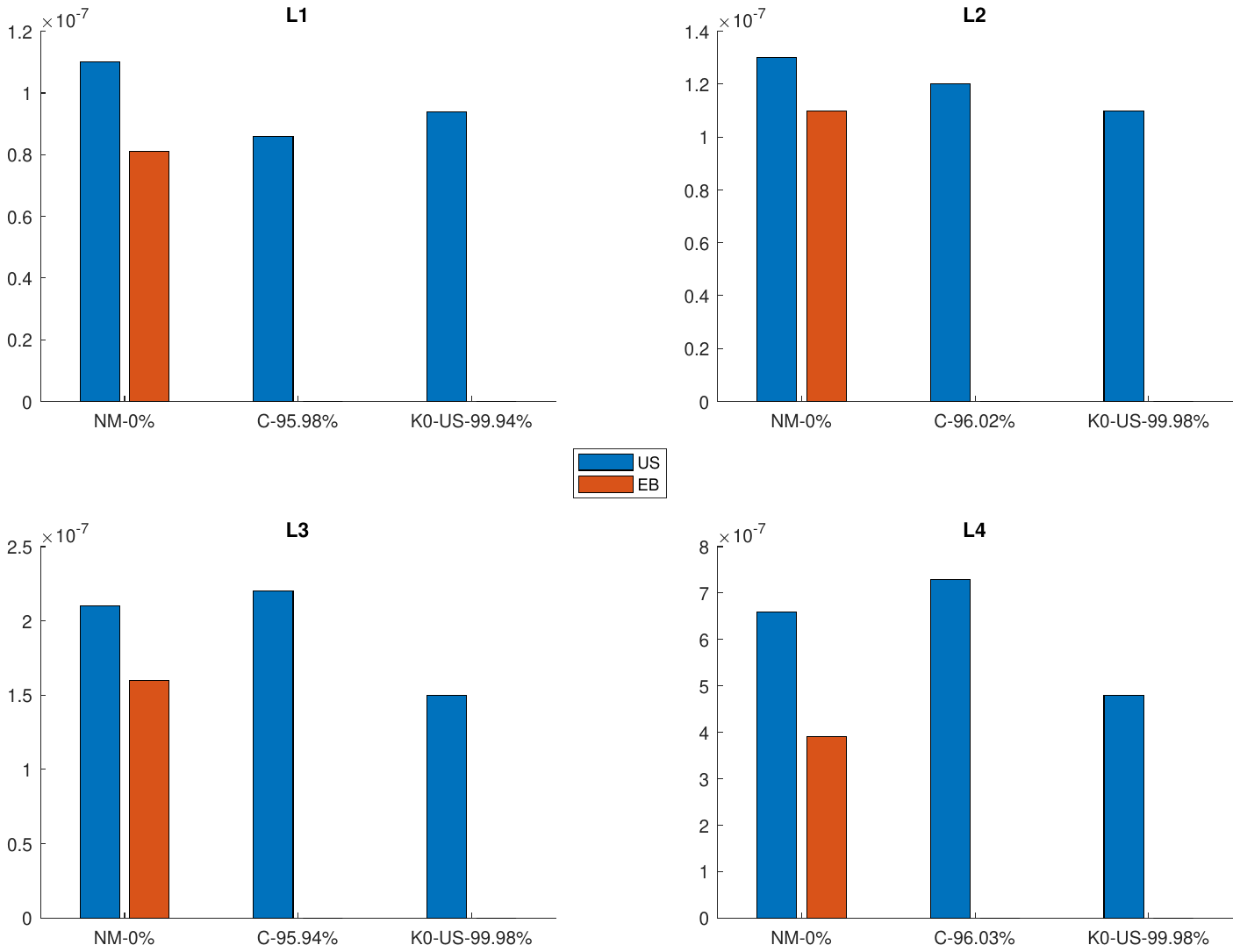}
	\caption{ {\bf Constant Space Models Query Times for the \textbf{amzn} Dataset }. For each memory level, the blue bar reports the average query time of Uniform Binary Search using, from left to right, no model, Cubic model and {\bf KO-US}. In addition, we report the average query time also for the Eytzinger Binary Search in the orange bar. 
	}\label{F:qamzn64}
\end{figure}

\begin{figure}[t]
	\centering
	\includegraphics[width=1\textwidth]{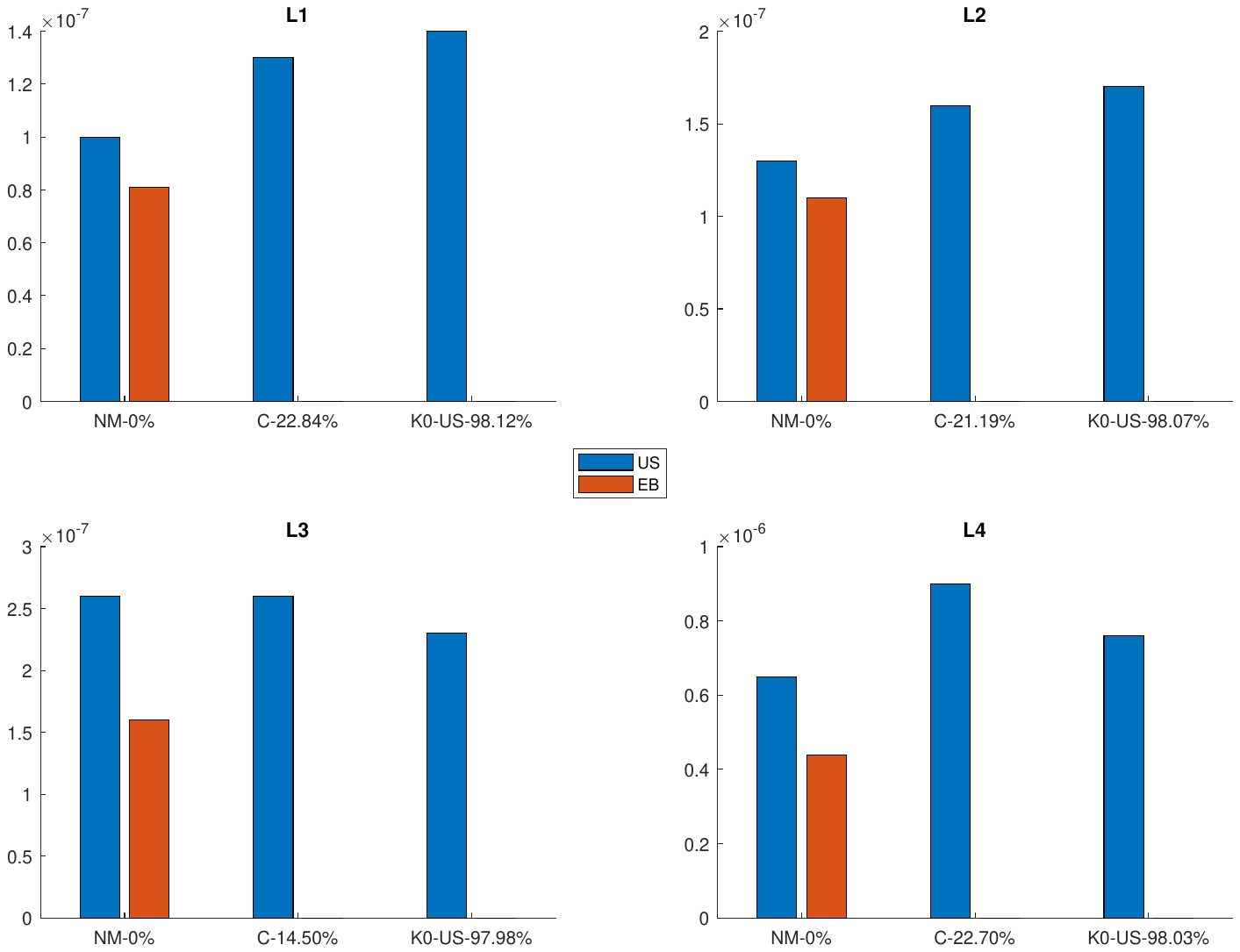}
	\caption{{\bf Constant Space Models Query Times for the \textbf{osm} Dataset.} The figure legend is as in Figure \ref{F:qamzn64}.}
	\label{F:qosm}
\end{figure}

The results of the experiments for this group of Models are reported in Figures \ref{F:qamzn64} and \ref{F:qosm} for the {\bf amzn} and the {\bf osm} datasets, respectively, and in Figures \ref{S-F:qface} and \ref{S-F:qwiki} of the Supplementary File, for the remaining ones.
In those figures, only the query time for  Uniform Binary Search is reported, since the results are analogous to the ones obtained by using the Standard routine. In addition, the query time for the Eytzinger Binary Search is also reported, as an useful baseline, because of its superiority among the classic routines that take constant additional space with respect to the size of the input table, as discussed in \cite{Morin17}. From the mentioned figures, it is evident that the query performance of each model considered here is highly influenced by how difficult to learn is the \emph{CDF} of the input table, as explained next.

\begin{itemize}

\item The Cubic Model achieves a high reduction factor, i.e. $\approx 99\%$, on the versions of the {\bf face } dataset for the first three levels of the internal memory hierarchy and it is also the best performing, even compared to the Eytzinger Layout routine. This is a quite remarkable achievement, but the involved datasets have an almost uniform  \emph{CDF}, while a few outliers disrupt such a uniformity on the {\bf L4} version of that dataset (see Figure \ref{fig:dataCDF} and the discussion regarding the {\bf face} dataset in Section \ref{sec:Datasets}).

\item The Learned $k$-ary Search Model achieves a high reduction factor on all versions of the {\bf amzn} and the {\bf wiki} datasets, i.e., $\approx 99.73$, and it is faster  than Uniform Binary Search and the
Cubic Model. Those datasets have a regular \emph{CDF} across all the internal memory levels. It is to be noted that the Eytzinger Layout routine is competitive with the  Learned $k$-ary Search Model. 

\item No constant space Learned Model wins on the difficult to learn dataset. The {\bf osm} dataset has a {\bf CDF} difficult to learn (see Figure \ref{fig:dataCDF}) and such a characteristic is preserved across the internal memory levels. The Learned $k$-ary Search Model achieves a quite respectable reduction factor, i.e., $\approx 98\%$, but no speed-up with respect to Uniform Binary Search. 
In order to get insights into such a counter-intuitive behaviour, we have performed an additional experiment. For each representative dataset and as far as the Learned $k$-ary Search Model  is concerned, we have computed  two kinds  of reduction factors: the first is the \vir{global} one, achieved considering the size of the entire table, while the second is the \vir{local} one, computed as the average among the reduction factors of each segment. Those results are reported in Table \ref{T:rf}. For the {\bf osm} dataset, it is evident that the \vir{local} reduction factors are consistently lower than the \vir{global} ones, highlighting that its \emph{CDF} is also locally difficult to approximate, which in turn implies an ineffective use of the local prediction for the Learned $k$-ary Search, resulting in poor time performance. Finally, it is to be noted that the 
Eytzinger Layout routine is the best performing.

\end{itemize}

In conclusion, in applications where there is a constant space constraint with respect to the input table and a layout other than sorted can be used, then the Eytzinger Binary Search is still the best choice, unless the \emph{CDF} of the input dataset is particularly easy to approximate. If such a layout cannot be afforded, the best choice is the use of a constant space Model, in particular the Learned $k$ary Search Model  only for datasets with a \emph{CDF} simple to approximate, otherwise the use of  Uniform Binary Search alone is indicated.

It is also of interest to pointed out that our research extends the results in \cite{Morin17} regarding the Eytzinger Binary Search routine: even compared to Learned Indexes that use constant space, it still results to be competitive and, many times superior to them.

\begin{table}
    \centering
    \caption{{\bf Global and Local Reduction Factors}. For the two representative datasets, i.e. {\bf amzn}  and {\bf osm}, and for each memory level, in each entry, we report the global reduction factor (left) and the local one (right).}    
    \label{T:rf}
    \begin{tabular}{|c|c|c|}
        \hline
         & amzn & osm \\ \hline
         L1 & 99.94 - 99.48 & 98.12 - 86.70 \\ \hline
         L2 & 99.98 - 99.56 & 98.07 - 86.57 \\ \hline
         L3 & 99.98 - 99.53 & 97.98 - 86.43 \\ \hline
         L4 & 99.98 - 99.54 & 98.03 - 86.57 \\ \hline
    \end{tabular}
\end{table}

\subsection{Parametric Space Models}\label{sec:qc-pm}
\begin{figure}[t]
	\centering
	\includegraphics[width=1\textwidth]{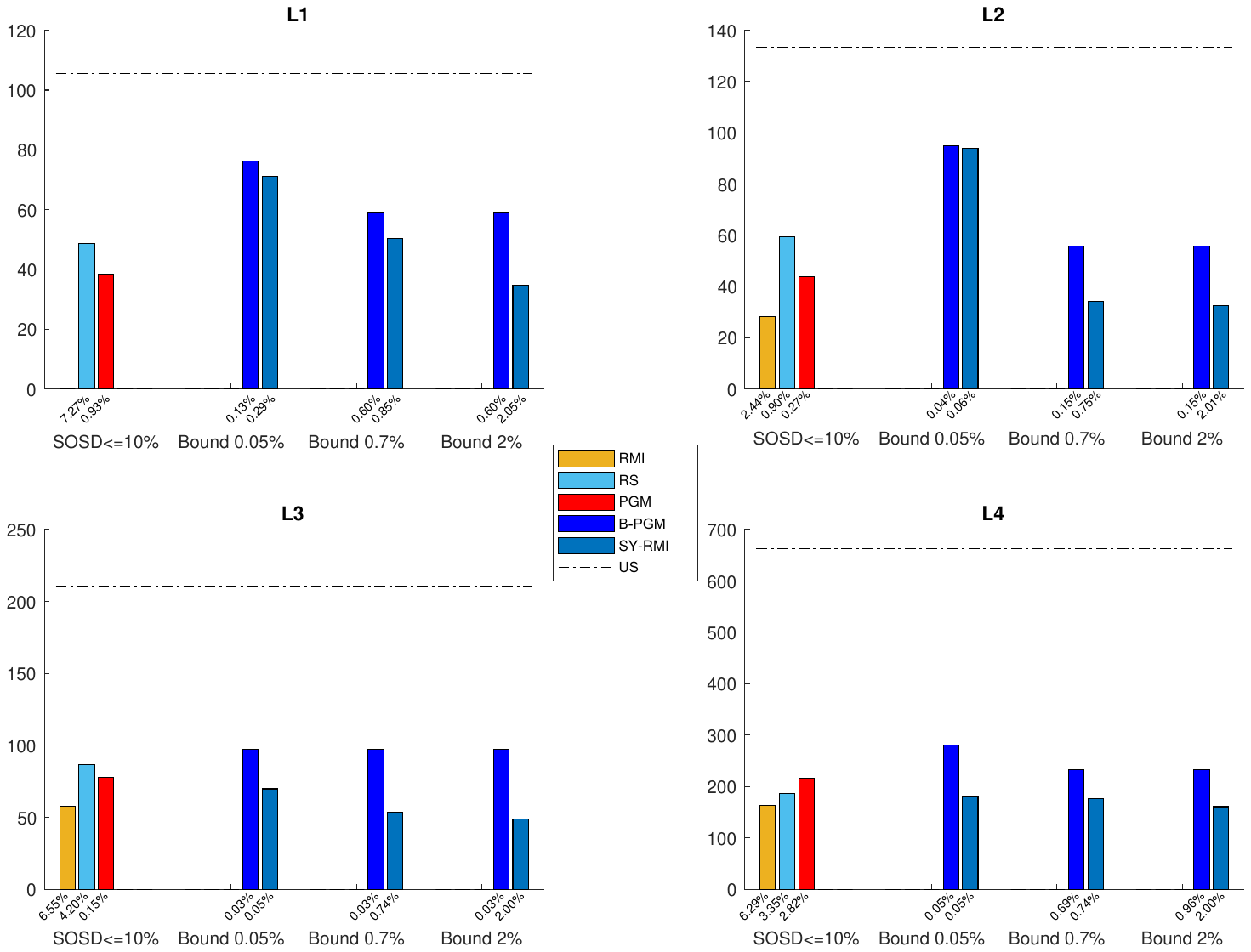}
	\caption{ {\bf Query times for the \textbf{amzn} dataset on Learned Indexes in Small Space.}  The methods are the ones in the legend (middle of the four panels, the notation is as in the main text and each method has a distinct colour). For each memory level, the abscissa reports methods grouped by space occupancy, as specified in the main text. When no model in a class output by {\bf SOSD} takes at most $10\%$ of additional space, that class is absent. The ordinate reports the average query time, with Uniform Binary Search executed in {\bf SOSD} as baseline (horizontal lines).}
	\label{F:sqamzn64n}
\end{figure}

\begin{figure}[t]
	\centering
	\includegraphics[width=1\textwidth]{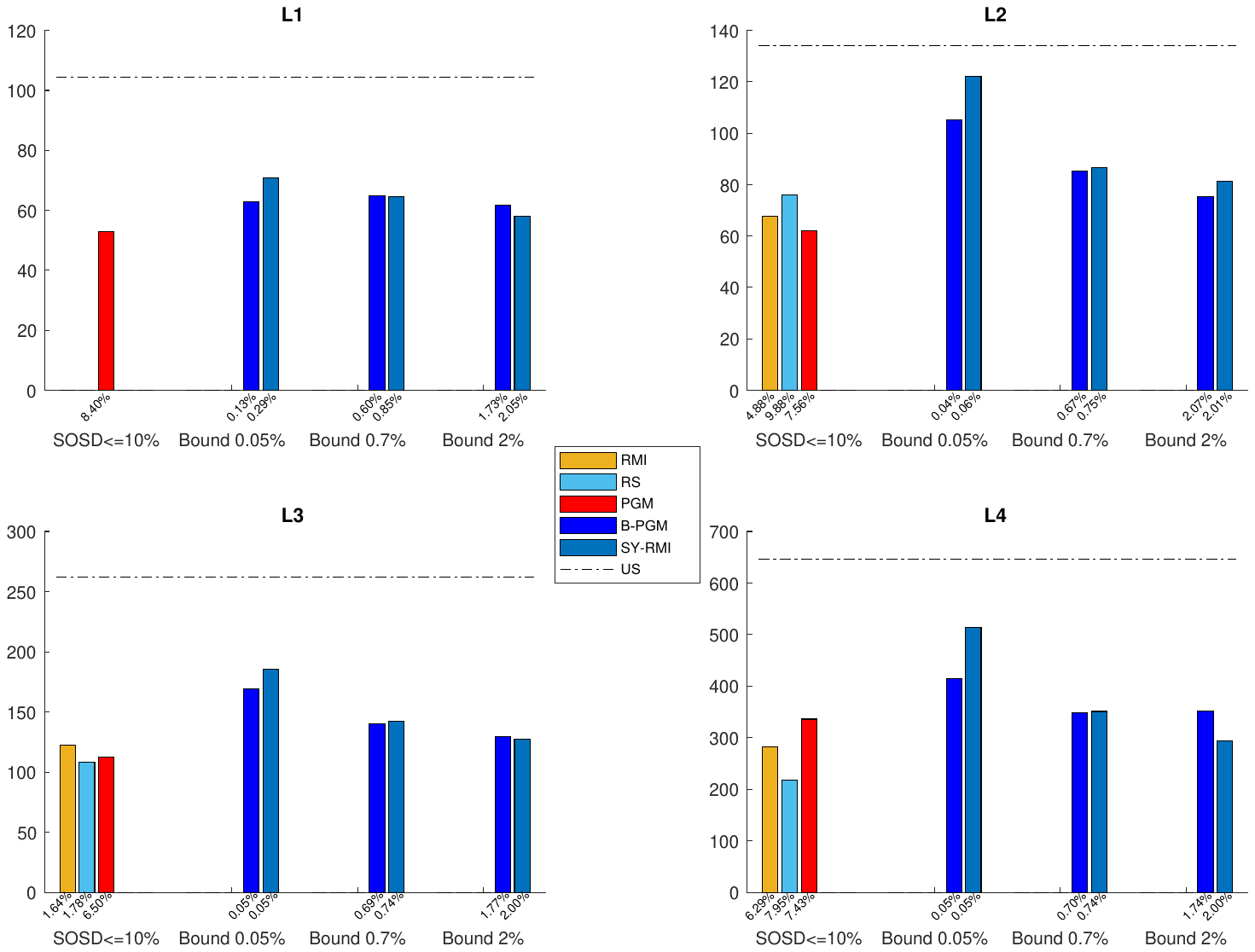}
	\caption{ {\bf Query times for the  \textbf{osm} dataset on Learned Indexes in Small Space.} The figure  legend is as the one in Figure   \ref{F:sqamzn64n}.  }
	\label{F:sqosmn}
\end{figure}

 For the convenience of the reader, we recall that the Model classes involved are: {\bf RMI}, {\bf RS}, {\bf PGM},  the Synoptic {\bf RMI} and the bi-criteria {\bf PGM}, which are trained on the input datasets (see Section \ref{sec:Datasets}), as reported in Section \ref{sec:LM}. The batch of queries used here are obtained as described in Section \ref{sec:Datasets}. For each of the first three Model classes, we consider, among the trained Models, the fastest in terms of query time and that takes less than $10\%$ of space, with respect to the one taken by the input table. For the other two Model classes, we consider three increasing bounds on space, i.e., $0.05\%$, $0.7\%$ and $2\%$, with respect to the space of the table alone, and take the average query time.  Moreover, as a measure of the Learned Indexes speed-up, we report also the query time of  Uniform Binary Search.  The results of the corresponding experiments are reported in Figures \ref{F:sqamzn64n} and \ref{F:sqosmn} for {\bf amzn} and {\bf osm} datasets, respectively, and in Figures \ref{S-F:sqfacen} and \ref{S-F:sqwikin} of the Supplementary File, for the remaining ones.

An interesting finding is that both the Synoptic {\bf RMI} and the bi-criteria {\bf PGM} perform better than Uniform Binary Search across datasets and memory levels using very little additional space. That is, one can enjoy the speed of Learned Indexes with a very small  space penalty. Moreover, it is important to note that, except for the {\bf L1} memory level, the space of those two Models is very close to the user-defined bound. Furthermore, in terms of query performances, such two Models seem to be complementary. In fact, the bi-criteria {\bf PGM} performs better on the {\bf L1} and {\bf L4} memory levels, while the Synoptic {\bf RMI} on the remaining ones. This complementary and good control of space make those two models quite useful in practice. 

In addition to those findings, our research provides some more insights into the relation time-space in Learned Indexes, extending the results in the benchmark study,  as  we now discuss.

\begin{itemize}
    \item {\bf Space  Constraints and  the  Models Provided by SOSD.}  We   have fixed  a rather  small   space budget,  i.e.,  at  most $10\%$ of additional space in order  for  a Model  returned  by {\bf SOSD} to be  considered. The {\bf RS} Index is not competitive with respect to the other Learned Indexes. Those latter consistently use less space and time, across datasets and memory levels. As for the {\bf RMI}s coming out of {\bf SOSD}, they are not able to operate in a small space at the {\bf L1} memory level. On the other memory levels, they are competitive with respect to the bi-criteria {\bf PGM} and the Synoptic {\bf RMI}, but they require more space with respect to them.

    \item {\bf Space, Time, Accuracy of Models.}  As stated in the benchmarking study, a common view of Learned Indexing Data Structures is as a \emph{CDF} lossy compressor, see also \cite{kraska18case,Ferragina:2020pgm}. In this view, the quality of a Learned Index can be judged by the size of the structure and its reduction factor. In that study, it is also argued that this view does not provide an accurate selection criterion for Learned Indexes. Indeed, it may very well be that an index structure with a excellent reduction factor takes a long time to produce a search bound, while an index structure with a worse reduction factor that quickly generates an accurate search bound may be of better use. In the benchmarking study, they also provide evidence that the space/time trade-off is the key factor in determining which Model to choose. Our contribution is to provide additional results supporting those findings. To this end, we have conducted several experiments, whose results are reported in Tables \ref{T:SyTAmzn}-\ref{T:SyTOsm} and Tables \ref{S-T:SyTFace}-\ref{S-T:SyTWiki} on the Supplementary File. In such Tables, for each dataset, we report a synopsis of three parameters, i.e., query time, space used in addition by the model and reduction factor, across all datasets and memory levels. In particular, for each dataset, we compare the best-performing model with all the ones that use small space, taking, for each parameter, the ratio Model/Best Model. The ratio values are reported from the second row of the table, the first row shows the average values of the parameters for the best model.  First, it is useful to note that, even in a small space model, it is possible to obtain a very well, if not nearly perfect, prediction (i.e. very high reduction factor). However, prediction power is somewhat marginal to assess performance. Indeed, across memory levels, we see a space classification of model configurations. The most striking feature of this classification is that the gain in query time between the best model and the others is within small constant factors, while the difference in space occupancy may be in most cases several orders of magnitude. That is, space is the key to efficiency.  
    
\end{itemize}

        \begin{table}[t]
            \begin{center}
                \caption{{\bf A Synoptic Table of Space, Time and Accuracy of Models on amzn Dataset}. For each memory level, we report in the first row the best performing method for that memory level. The columns named time, space and reduction factor indicate for this best model, the average query time in seconds, the average additional space used in Kb and the average of the empirical reduction factor. From the second row, we report the versions of the {\bf RMI}, {\bf RS}, {\bf PGM} and Synoptic {\bf RMI} models that use the least space. In particular, the number next to the Models represent in percentage the bound on the used space with respect to the input dataset. The columns indicate now the ratio Model/best Model of the time, space and reduction factor.}\label{T:SyTAmzn}
                \scriptsize

        \begin{tabular}{|c|c|c|c|}
        \hline \multicolumn{4}{|c|}{L1}  \\ \hline \hline
        & Time & Space & Reduction Factor \\ \hline
        
		Best RMI & 1.89e+01 & 3.09e+00 & 99.84 \\ \hline\hline
		B-PGM 0.05 & 4.03e+00 & 1.30e-02 & 2.50e-01 \\ \hline
		SY-RMI 0.05 & 3.77e+00 & 2.85e-02 & 1.75e-01 \\ \hline
		RS $<$ 10 & 2.58e+00 & 7.06e-01 & 9.29e-01 \\ \hline
		Best RMI & 1.00e+00 & 1.00e+00 & 1.00e+00 \\ \hline

        \hline \multicolumn{4}{|c|}{L2}  \\ \hline \hline
        & Time & Space & Reduction Factor \\ \hline
        
		Best RMI & 2.51e+01 & 6.16e+00 & 99.97 \\ \hline\hline
		B-PGM 0.05 & 3.78e+00 & 1.62e-02 & 9.16e-01 \\ \hline
		SY-RMI 0.05 & 3.74e+00 & 2.60e-02 & 6.44e-01 \\ \hline
		Best RS $<$ 10 & 2.38e+00 & 3.68e-01 & 9.92e-01 \\ \hline
		Best RMI & 1.00e+00 & 1.00e+00 & 1.00e+00 \\ \hline
		\end{tabular}
		\qquad
        \begin{tabular}{|c|c|c|c|}
        \hline \multicolumn{4}{|c|}{L3}  \\ \hline \hline
        & Time & Space & Reduction Factor \\ \hline
        
		Best RMI & 4.70e+01 & 6.29e+03 & 100.00 \\ \hline\hline
		B-PGM 0.05 & 2.07e+00 & 3.05e-04 & 1.00e+00 \\ \hline
		SY-RMI 0.05 & 1.49e+00 & 4.79e-04 & 9.99e-01 \\ \hline
		RS $<$ 10 & 1.59e+00 & 4.00e-02 & 1.00e+00 \\ \hline
		RMI $<$ 10 & 1.03e+00 & 6.25e-02 & 1.00e+00 \\ \hline

        \hline \multicolumn{4}{|c|}{L4}  \\ \hline \hline
        & Time & Space & Reduction Factor \\ \hline
        
		Best RMI & 1.51e-07 & 2.01e+05 & 100.00 \\ \hline\hline
		B-PGM 0.05 & 1.85e+00 & 3.93e-03 & 1.00e+00 \\ \hline
		SY-RMI 0.05 & 1.18e+00 & 3.97e-03 & 1.00e+00 \\ \hline
		Best RS & 1.19e+00 & 7.16e-02 & 1.00e+00 \\ \hline
		RMI $<$ 10 & 1.03e+00 & 5.00e-01 & 1.00e+00 \\ \hline
		\end{tabular}	
		\end{center}
	\end{table}	
	
	\begin{table}[t]
            \begin{center}
                \caption{{\bf A Synoptic Table of Space, Time and Accuracy of Models on osm Dataset}. The legend is as in \ref{T:SyTAmzn}}\label{T:SyTOsm}
                \scriptsize

        \begin{tabular}{|c|c|c|c|}
                \hline \multicolumn{4}{|c|}{L1}  \\ \hline \hline
                & Time & Space & Reduction Factor \\ \hline
        
		Best RMI & 2.72e+01 & 1.15e+03 & 99.87 \\ \hline\hline
		B-PGM 0.05 & 2.31e+00 & 3.49e-05 & 1.74e-01 \\ \hline
		SY-RMI 0.05 & 2.60e+00 & 7.67e-05 & 2.30e-01 \\ \hline
		Best RMI & 1.00e+00 & 1.00e+00 & 1.00e+00 \\ \hline
		Best RS & 1.19e+00 & 4.33e+01 & 9.99e-01 \\ \hline

        \hline \multicolumn{4}{|c|}{L2}  \\ \hline \hline
        & Time & Space & Reduction Factor \\ \hline
        
		Best RMI & 3.93e+01 & 1.84e+03 & 99.97 \\ \hline\hline
		B-PGM 0.05 & 2.68e+00 & 5.45e-05 & 7.75e-01 \\ \hline
		SY-RMI 0.05 & 3.11e+00 & 8.72e-05 & 7.24e-01 \\ \hline
		RMI $<$ 10 & 1.73e+00 & 6.71e-03 & 9.87e-01 \\ \hline
		RS $<$ 10 & 1.93e+00 & 1.36e-02 & 9.79e-01 \\ \hline
		\end{tabular}
		\qquad
        \begin{tabular}{|c|c|c|c|}
                \hline \multicolumn{4}{|c|}{L3}  \\ \hline \hline
                & Time & Space & Reduction Factor \\ \hline
        
		Best RS & 7.06e+01 & 4.63e+04 & 100.00 \\ \hline\hline
		B-PGM 0.05 & 2.40e+00 & 6.22e-05 & 9.98e-01 \\ \hline
		SY-RMI 0.05 & 2.63e+00 & 6.52e-05 & 9.31e-01 \\ \hline
		RMI $<$ 10 & 1.75e+00 & 2.12e-03 & 9.97e-01 \\ \hline
		RS $<$ 10 & 1.55e+00 & 2.31e-03 & 1.00e+00 \\ \hline
		\hline
		\hline \multicolumn{4}{|c|}{L4}  \\ \hline \hline
        & Time & Space & Reduction Factor \\ \hline
        
		Best RS & 2.04e-07 & 5.08e+05 & 100.00 \\ \hline\hline
		SY-RMI 0.05 & 2.52e+00 & 1.57e-03 & 9.99e-01 \\ \hline
		B-PGM 0.05 & 2.03e+00 & 1.59e-03 & 1.00e+00 \\ \hline
		RMI $<$ 10 & 1.18e+00 & 1.98e-01 & 1.00e+00 \\ \hline
		RS $<$ 10 & 1.05e+00 & 2.50e-01 & 1.00e+00 \\ \hline
		\end{tabular}	
		\end{center}
	\end{table}

\section{Conclusions and Future Directions}\label{sec:concl}
In this research, we have provided a systematic experimental analysis regarding the ability of Learned Model Indexes to perform better than Binary Search in small space. 
Although not as simple as it seems, we show that this is indeed possible. However, our results also indicate that there is  a big gap between the best performing methods and the others we have considered and that operate in small space. Indeed, the query time performance of the latter with respect to the former is bounded by small constants, while the space usage may differ even by five orders of magnitude. 
This brings to light the acute need to investigate the existence of \vir{small space} models that should close the time gap mentioned earlier. 
Another important aspect, with potential practical impact, is to devise models that can work on layouts other than Sorted, i.e., Eytzinger. Finally, given that Eytzinger Binary Search is consistently faster than Binary Search for datasets fitting in main memory, an investigation of how different variants of Binary Search perform in {\bf SOSD}, using both new and State of the Art models, also deserves to be investigated.

\section{Acknowledgement}

The authors are deeply indebted to the Associate Editor and the two Referees, for the very stimulating comments received during the entire revision process.
This research is funded in part by MIUR Project of National Relevance 2017WR7SHH “Multicriteria Data Structures and Algorithms: from compressed to learned indexes, and beyond”. Additional support to RG has been granted by Project INdAM - GNCS  “Modellizazzione ed analisi di big knowledge graphs per la risoluzione di problemi in ambito medico e web”.

\bibliographystyle{plain}
\bibliography{references.bib}

\end{document}


\title{Learned Sorted Table Search and Static Indexes in Small Model Space-Supplementary  File	\footnote{This research is funded in part by MIUR Project of National Relevance 2017WR7SHH “Multicriteria Data Structures and Algorithms: from compressed to learned indexes, and beyond”. We also acknowledge an NVIDIA Higher Education and Research Grant (donation of a Titan V GPU). Additional support to RG has been granted by Project INdAM - GNCS  “Analysis and Processing of Big Data based on Graph Models” }}
	
	\author{Domenico Amato$^1$\\
		\and
		Giosu\'e Lo Bosco$^1$\footnote{corresponding author, email: giosue.lobosco@unipa.it}\\
		\and
		Raffaele Giancarlo$^1$}

	\date{
		$^1$Dipartimento di Matematica e Informatica\\ 
		Universit\'a degli studi di Palermo, ITALY\\
		\today
	}

	\maketitle
	
	\begin{abstract}
		This document provides additional details with respect to the content of the Main Manuscript by the same title.
	\end{abstract}

	\section{Binary Search and Its Variants}\label{sec:BS}
	
    With reference to the routines mentioned in the Main Text (Section \ref{M-sec:methods}) and following research in \cite{Morin17}, we provide more details about two kind of layouts.

\begin{figure}[tbh]
		\begin{center}
			\includegraphics[width=0.40\textwidth]{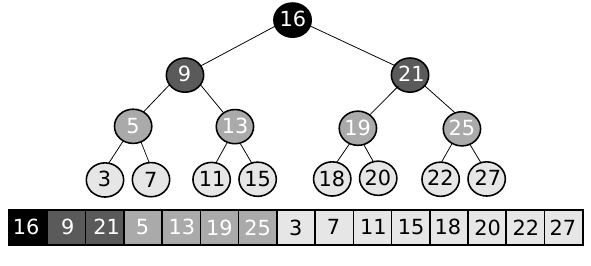}
			\caption{{\bf An example of  Eyzinger layout of a table with 15 elements. } See also \cite{Morin17}.}
			\label{fig:EyL}
		\end{center}
	\end{figure}

	\begin{itemize}
		
		\item [1] Sorted. We use two version of Binary Search for this layout. The template of the {\bf lower\_bound} routine is provide in  Algorithm \ref{AL:U-LB}, while the Uniform Binary Search implementation is given in Algorithm \ref{AL:BFS}. In particular, such an implementation of Binary Search is as in \cite{Morin17}.
	
		\item [2] Eytzinger Layout \cite{Morin17}.  The sorted table is now seen as stored in a virtual complete balanced binary search tree. Such a tree is laid out in Breadth-First Search order in an array. An example is provided in Fig. \ref{fig:EyL}. The implementation is reported in Algorithm \ref{AL:BFE}. 
	\end{itemize}
	
	\begin{algorithm}
    	\caption{{\bf lower\_bound Template}.}
    	\label{AL:U-LB}
    	\begin{algorithmic}[1]
    		\BState ForwardIterator lower\_bound (ForwardIterator first, ForwardIterator last, const T\& val)\{
    		\State \ \	ForwardIterator it;
    		\State \ \	iterator\_traits$<$ForwardIterator$>$::difference\_type count, step;
    		\State \ \	count = distance(first,last);
    		\State \ \	while (count$>$0)\{
    		\State \ \ \ \		it = first; step=count/2; advance (it,step);
    		\State \ \ \ \		if (*it$<$val)\{
    		\State \ \ \ \ \ \			first=++it;
    		\State \ \ \ \ \ \			count-=step+1;
    		\State \ \ \ \		\}
    		\State \ \	\ \ 	else count=step;
    		\State \ \	\}
    		\State \ \	return first;
    		\BState \}
    	\end{algorithmic}
    \end{algorithm}

	\begin{algorithm}
		\caption{{\bf Implementation of Uniform Binary Search} The code is as in \cite{Morin17} (see also \cite{KnuthS,Shutz18} ).}
		\label{AL:BFS}
		\begin{algorithmic}[1]
			\BState int UniformBinarySearch(int *A, int x,  int left, int right)\{
			\State \ \ \ const int *base = A;
			\State \ \ \ int n = right;
			\State \ \ \ while (n $>$ 1) \{
			\State	\ \ \ \ \ const int half = n / 2;
			\State	\ \ \ \ \ \_\_builtin\_prefetch(base + half/2, 0, 0);
			\State	\ \ \ \ \ \_\_builtin\_prefetch(base + half + half/2, 0, 0);
			\State	\ \ \ \ \ base = (base[half] $<$ x) ? \&base[half] : base;
			\State	\ \ \ \ \ n -= half;
			\State	\ \ \ \}
			\State \ \ \ return (*base $<$ x) + base - A;
			\BState \}	
		\end{algorithmic}
	\end{algorithm}

	\begin{algorithm}
		\caption{{\bf  Uniform Binary Search with Eytzinger layout} The code is as in   \cite{Morin17}.}
		\label{AL:BFE}
		\begin{algorithmic}[3]
			\BState int EytzingerLayoutSearch(int *A, int x,  int left, int right)\{
			\State \ \ \ int i = 0;
			\State \ \ \ int n = right;
			\State \ \ \ while (i $<$ n)\{
			\State \ \ \ \ \ \_\_builtin\_prefetch(A+(multiplier*i + offset)); 
			\State \ \ \ \ \ i = (x $<=$ A[i]) ? (2*i + 1) : (2*i + 2);
			\State \ \ \ \}
			\State \ \ \ int j = (i+1) $>>$ \_\_builtin\_ffs($\sim$(i+1));
			\State \ \ \ return (j == 0) ? n : j-1;
			\BState \}	
		\end{algorithmic}
	\end{algorithm}
	
	\section{Training of the Novel Models: Analysis and Insights into Model Training - Additional Results}\label{Sec:LM}
	
	Following the same approach used in Section \ref{M-sec:LM} of the Main Text, we divide the training time analysis into two groups: Constant and Parametric Space Models.
	
	\begin{itemize}
	    \item Tables \ref{T:CMTTL1}-\ref{T:CMTTL3} report the experiments concerning the Constant Space Models for the datasets L1, L2 and L3. 
	    \item Tables \ref{T:PMTTL1}-\ref{T:PMTTL3} report the experiments concerning the Parametric Space Models for the datasets L1, L2 and L3. 
	\end{itemize}

    \begin{table}
	    \begin{center}
    		\caption{ {\bf Constant Space Models Training time for {\bf L1} Tables}. The first column indicated the datasets. The remaining columns indicate the model used for the learning phase. Each entry reports the training time in seconds and per element.}\label{T:CMTTL1}
    		\small
    		\begin{tabular}{|c|c|c|}
    		    \hline
    			 &  {\bf KO-US} & {\bf C} \\
    			\hline
			
				{\bf amzn}  &5.3e-07&1.0e-07\\
				\hline
				{\bf face} &5.5e-07&8.5e-08\\
				\hline
				{\bf osm}  &4.6e-07&9.9e-08\\
				\hline
				{\bf wiki}  &9.0e-07&7.9e-08\\
    			\hline
    		\end{tabular}
    	\end{center}
    \end{table}
    
    \begin{table}
		\begin{center}
			\caption{{\bf Constant Space Models Training Time for {\bf L2} Tables}. The table legend is as in Table  \ref{T:CMTTL1}}\label{T:CMTTL2}
			\small
			\begin{tabular}{|c|c|c|}
			\hline
			   
			    & {\bf KO-BFS} & {\bf C} \\
			    \hline
				{\bf amzn} &1.8e-07&3.1e-07\\
				\hline
				{\bf face} &1.0e-07&2.8e-07\\
				\hline
				{\bf osm} &1.2e-07&2.9e-07\\
				\hline
				{\bf wiki} &1.0e-07&2.7e-07\\
				\hline
			\end{tabular}
		\end{center}
	\end{table}
    
    \begin{table}
	\begin{center}
		\caption{ {\bf Constant Space Models Training time for {\bf L3} Tables}. The table legend is as in Table  \ref{T:CMTTL1}.}\label{T:CMTTL3}
		\small
		\begin{tabular}{|c|c|c|}
		    \hline
		     
			 &  {\bf KO-US} & {\bf C} \\
			\hline
			
				{\bf amzn}  &6.3e-08&1.9e-08\\
				\hline
				{\bf face}  &3.9e-08&1.9e-08\\
				\hline
				{\bf osm}  &4.4e-08&2.0e-08\\
				\hline
				{\bf wiki}  &4.1e-08&1.9e-08\\
			\hline
		\end{tabular}
	\end{center}
\end{table}
    
    \begin{table}
	\begin{center}
		\caption{ {\bf Parametric Space Models Training Time for {\bf L1} Tables}. The first column indicated the datasets. The remaining columns indicate the model used for the learning phase. In particular, each entry reports the time used by {\bf CDFShop} and {\bf SOSD} libraries to train the entire batch of parametric models in seconds and per element.}\label{T:PMTTL1}
    		\small
    		\begin{tabular}{|c|c|c|c|c|}
		    \hline
		     
			 & {\bf CDFShop SY-RMI 2\%} &  {\bf CDFShop RMI}& {\bf SOSD RS} & {\bf SOSD PGM}\\
			\hline
			
				{\bf amzn} &5.2e-06  &5.6e-06 &3.5e-06 &5.0e-07\\
				\hline
				{\bf face}  &4.1e-06 &4.6e-06 &1.1e-06 &3.9e-07\\
				\hline
				{\bf osm}  &2.8e-04&2.9e-04 &6.9e-06 &4.0e-07\\
				\hline
				{\bf wiki} &7.8e-06&9.3e-06 &1.0e-05 &3.7e-07\\
    			\hline
    		\end{tabular}
    	\end{center}
    \end{table}

	\begin{table}
		\begin{center}
			\caption{{\bf Parametric Space Models Training time for {\bf L2} Tables}. The table legend is as in Table  \ref{T:PMTTL1}}\label{T:PMTTL2}
			\small
			\begin{tabular}{|c|c|c|c|c|}
			\hline
			   
			    &  {\bf CDFShop SY-RMI 2\%} & {\bf CDFShop RMI} & {\bf SOSD RS} & {\bf SOSD PGM}\\
				\hline
				{\bf amzn} &5.2e-06&5.6e-07 &3.5e-07 &5.0e-08\\
				\hline
				{\bf face} &4.1e-06&4.6e-07 &1.1e-07 &3.9e-08\\
				\hline
				{\bf osm} &2.8e-04&2.9e-05 &6.9e-07 &4.0e-08\\
				\hline
				{\bf wiki} &7.8e-06&9.3e-07 &1.0e-06 &3.7e-08\\
				\hline
			\end{tabular}
		\end{center}
	\end{table}

\begin{table}
	\begin{center}
		\caption{ {\bf Parametric Space Models Training time for {\bf L3} Tables}. The table legend is as in Table  \ref{T:PMTTL1}}\label{T:PMTTL3}
		\small
		\begin{tabular}{|c|c|c|c|c|}
		    \hline
		     
			 & {\bf CDFShop SY-RMI 2\%} &  {\bf CDFShop RMI}& {\bf SOSD RS} & {\bf SOSD PGM}\\
			\hline
			
				{\bf amzn}  &1.5e-06 &1.3e-07 &2.4e-08 &3.4e-08\\
				\hline
				{\bf face}  &1.5e-05&1.6e-06 &1.4e-08 &2.4e-08\\
				\hline
				{\bf osm}  &1.2e-05 &1.3e-06 &3.5e-08 &3.8e-08\\
				\hline
				{\bf wiki}  &2.3e-06&2.2e-07 &5.1e-08 &3.7e-08\\
			\hline
		\end{tabular}
	\end{center}
\end{table}

	\section{Query Experiments - Additional Results}
	
	In this Section, we report the experiments described and discussed in Section \ref{M-sec:qc} of the Main Text for the {\bf face} and {\bf wiki} datasets.
	
\begin{figure}[t]
		\centering
		\includegraphics[width=1\textwidth]{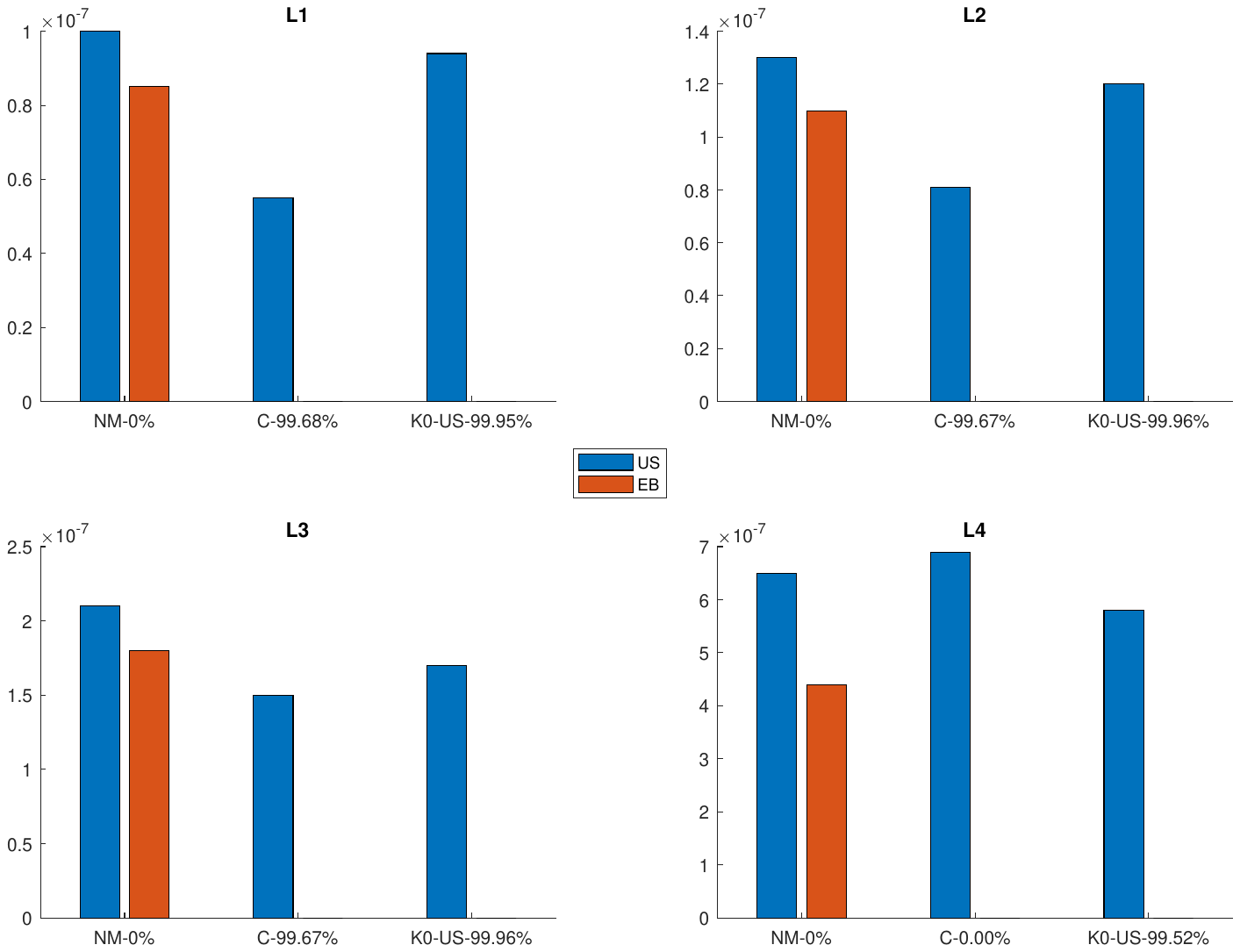}
		\caption{{\bf Constant Space Models Query Times for the \textbf{face} Dataset }. For each memory level, the blue bar reports the average query time of Uniform Binary Search using, from left to right, no model, Cubic model and {\bf KO-US}. In addition, we report the average query time also for the Eytzinger Binary Search in the orange bar.}\label{F:qface}
	\end{figure}

\begin{figure}[t]
		\centering
		\includegraphics[width=1\textwidth]{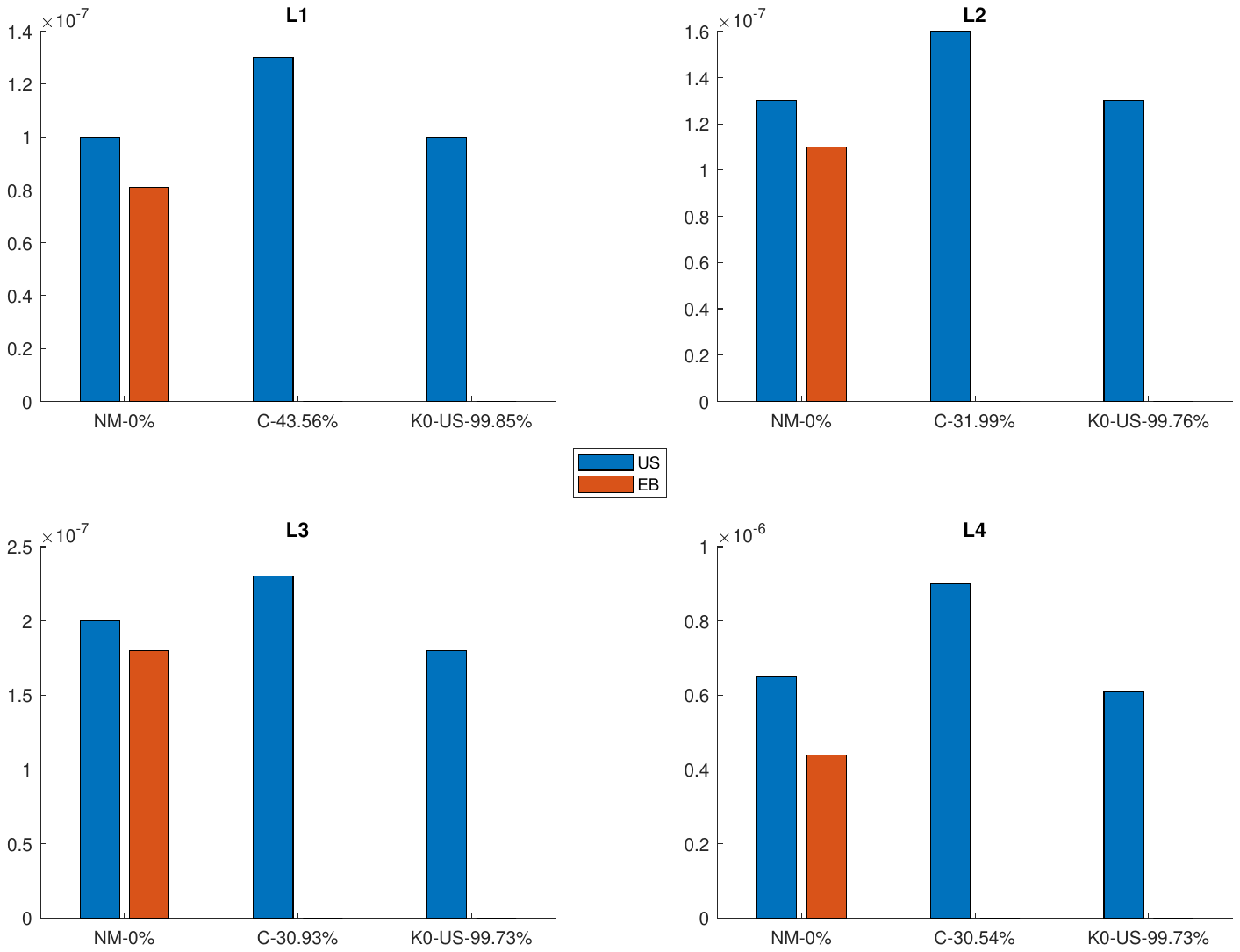}
		\caption{{\bf Constant Space Models Query Times for the \textbf{wiki} Dataset }. The legend is as in Figure \ref{F:qface} }
		\label{F:qwiki}
	\end{figure}

\begin{figure}[t]
		\centering
		\includegraphics[width=1\textwidth]{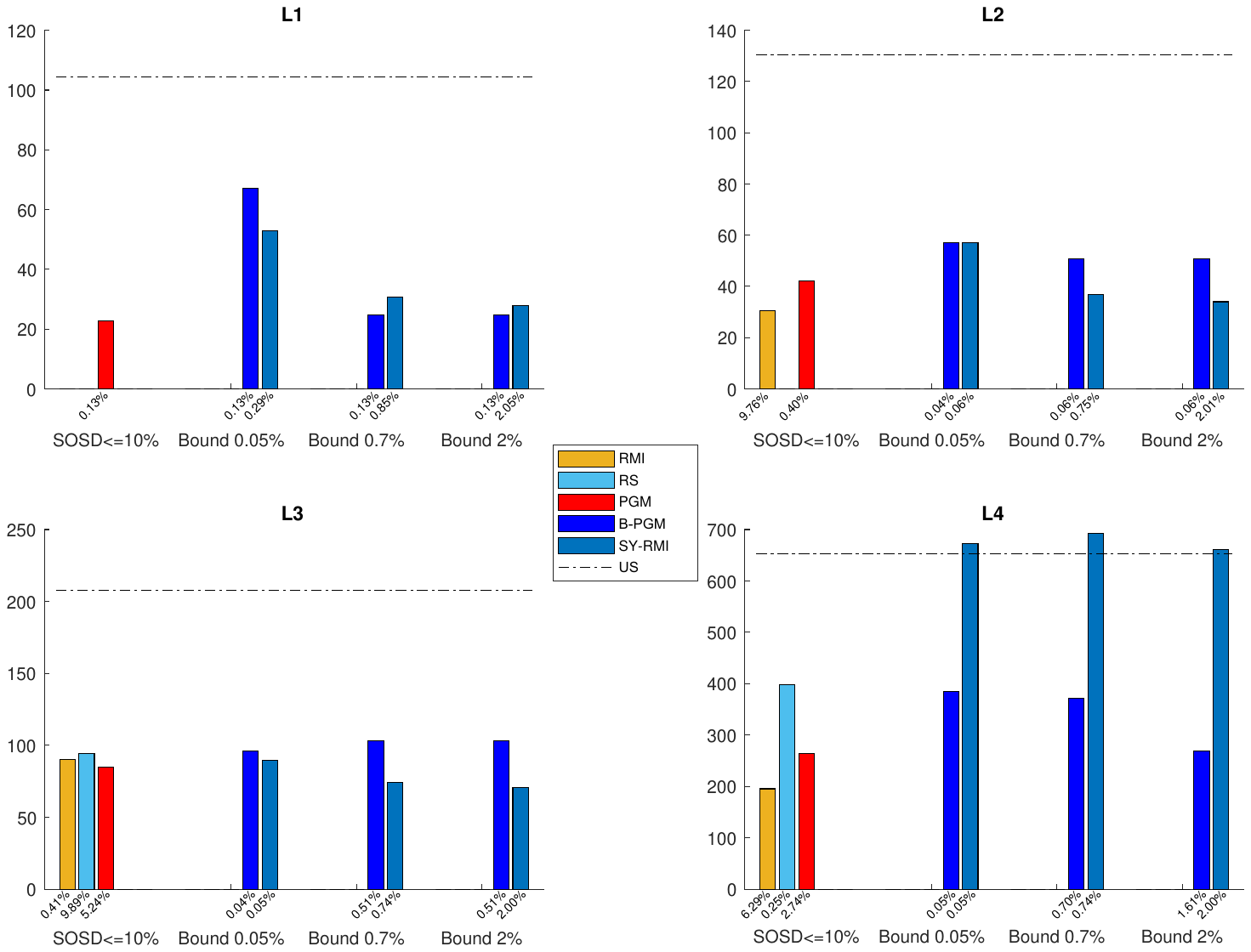}
		\caption{ {\bf Query times for the \textbf{face} dataset on Learned Indexes in Small Space.} The methods are the ones in the legend (middle of the four panels, the notation is as in the main text and each method has a distinct colour). For each memory level, the abscissa reports methods grouped by space occupancy, as specified in the main text. When no model in a class output by {\bf SOSD} takes at most $10\%$ of additional space, that class is absent. The ordinate reports the average query time, with Uniform Binary Search executed in {\bf SOSD} as baseline (horizontal lines). }
		\label{F:sqfacen}
		
	\end{figure}

\begin{figure}[t]
		\centering
		\includegraphics[width=1\textwidth]{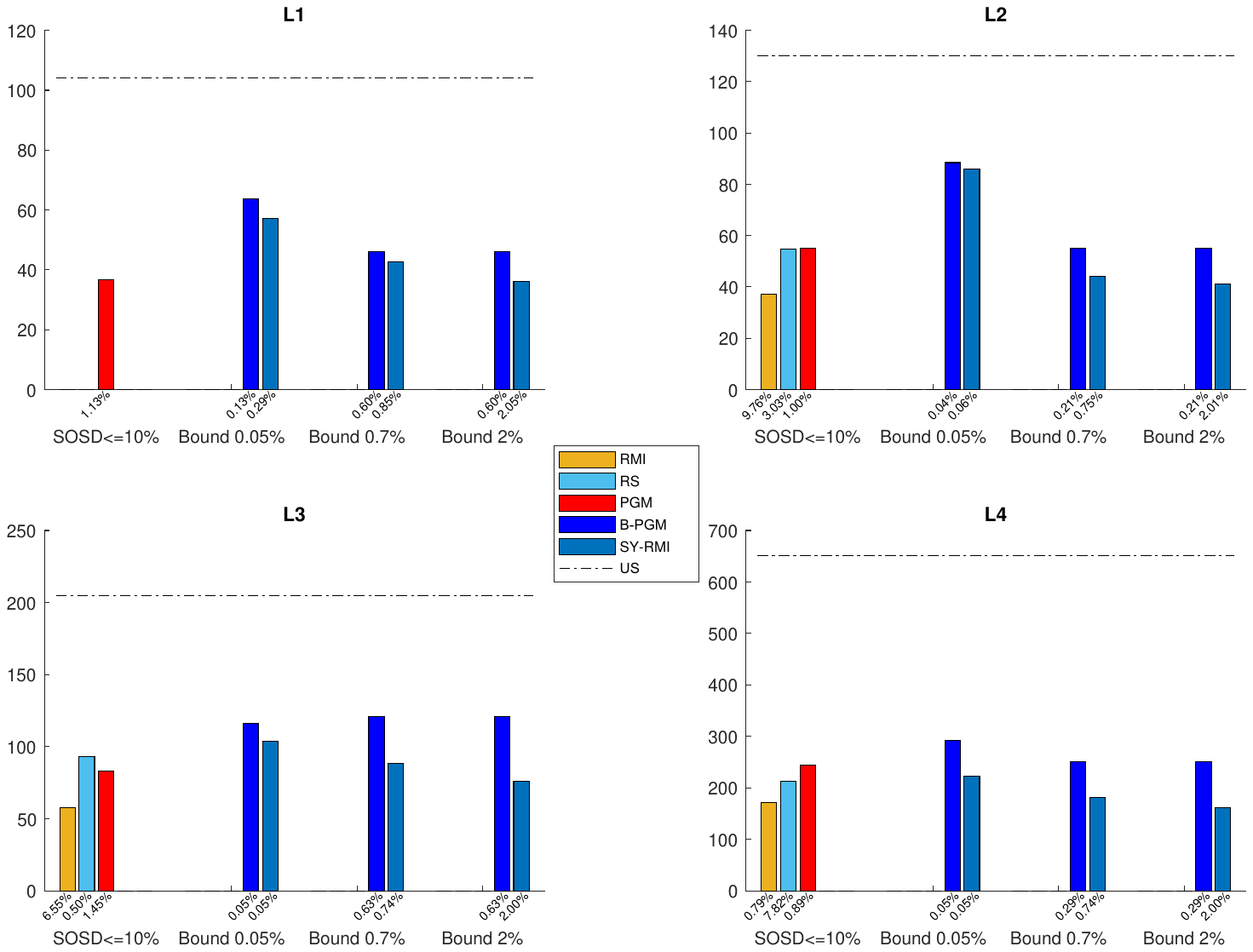}
		\caption{ {\bf Query times for the  \textbf{wiki} dataset on Learned Indexes in Small Space.} The figure  legend is as in Figure   \ref{F:sqfacen}.}
		\label{F:sqwikin}
		
	\end{figure}

	\begin{itemize}
	    \item Figures  \ref{F:qface}-\ref{F:qwiki} report the experiments concerning the Constant Space Models, as in Section \ref{M-sec:qc-cm}.
	    \item Figures  \ref{F:sqfacen}-\ref{F:sqwikin} report the experiments concerning the Parametric Space Models, as in Section \ref{M-sec:qc-pm}
	    \item  Tables \ref{T:SyTFace}-\ref{T:SyTWiki} report a synopsis of three parameters, i.e., query time, space used in addition by the model and reduction factor, as described in Section \ref{M-sec:qc-pm}.
	\end{itemize}

        \begin{table}[t]
            \begin{center}
                \caption{{\bf A Synoptic Table of Space, Time and Accuracy of Models on face Dataset}. For each memory level, we report in the first row the best performing method for that memory level. The columns named time, space and reduction factor indicate for this best model, the average query time in seconds, the average additional space used and the average of the empirical reduction factor. From the second row, we report the versions of the {\bf RMI}, {\bf RS}, {\bf PGM} and Synoptic {\bf RMI} models that use the least space. In particular, the number next to the Models represent the percentage of the used space with respect to the input dataset. The columns indicate now the ratio Model/best Model of the time, space and reduction factor.}\label{T:SyTFace}
                \scriptsize

        \begin{tabular}{|c|c|c|c|}
                \hline \multicolumn{4}{|c|}{L1}  \\ \hline \hline
                & Time & Space & Reduction Factor \\ \hline
        
		Best PGM & 2.26e+01 & 4.00e-02 & 99.52 \\ \hline\hline
		Best PGM & 1.00e+00 & 1.00e+00 & 1.00e+00 \\ \hline
		SY-RMI 0.05 & 2.33e+00 & 2.20e+00 & 6.30e-01 \\ \hline
		Best RMI & 1.16e+00 & 7.72e+01 & 1.00e+00 \\ \hline
		Best RS & 1.17e+00 & 5.90e+04 & 1.00e+00 \\ 
		\hline
        \hline \multicolumn{4}{|c|}{L2}  \\ \hline \hline
        & Time & Space & Reduction Factor \\ \hline
        
		Best RMI & 3.02e+01 & 1.23e+01 & 99.98 \\ \hline\hline
		B-PGM 0.05 & 1.89e+00 & 8.13e-03 & 9.98e-01 \\ \hline
		SY-RMI 0.05 & 1.89e+00 & 1.30e-02 & 9.40e-01 \\ \hline
		Best RMI & 1.00e+00 & 1.00e+00 & 1.00e+00 \\ \hline
		Best RS & 1.10e+00 & 1.92e+02 & 1.00e+00 \\ \hline\end{tabular}\qquad
        \begin{tabular}{|c|c|c|c|}
        \hline \multicolumn{4}{|c|}{L3}  \\ \hline \hline
        & Time & Space & Reduction Factor \\ \hline
        
		Best RMI & 6.11e+01 & 7.86e+02 & 100.00 \\ \hline\hline
		B-PGM 0.05 & 1.57e+00 & 3.33e-03 & 1.00e+00 \\ \hline
		RMI $<$ 10 & 1.19e+00 & 3.13e-02 & 1.00e+00 \\ \hline
		SY-RMI 0.7 & 1.22e+00 & 5.62e-02 & 1.00e+00 \\ \hline
		RS $<$ 10 & 1.53e+00 & 7.54e-01 & 1.00e+00 \\ \hline
        \hline \multicolumn{4}{|c|}{L4}  \\ \hline \hline
        & Time & Space & Reduction Factor \\ \hline
        
		Best RMI & 1.80e-07 & 2.01e+05 & 100.00 \\ \hline\hline
		SY-RMI 0.05 & 3.74e+00 & 3.97e-03 & 1.32e-02 \\ \hline
		B-PGM 0.05 & 2.14e+00 & 3.98e-03 & 1.00e+00 \\ \hline
		Best RS & 2.21e+00 & 3.96e-02 & 1.00e+00 \\ \hline
		RMI $<$ 10 & 1.06e+00 & 5.00e-01 & 1.00e+00 \\ \hline
		\end{tabular}
		\end{center}
	\end{table}

        \begin{table}[t]
            \begin{center}
                \caption{{\bf A Synoptic Table of Space, Time and Accuracy of Models on wiki Dataset}. The legend is as in \ref{T:SyTFace}.}\label{T:SyTWiki}
                \scriptsize

        \begin{tabular}{|c|c|c|c|}
                \hline \multicolumn{4}{|c|}{L1}  \\ \hline \hline
                & Time & Space & Reduction Factor \\ \hline
        
		Best RMI & 2.55e+01 & 3.09e+00 & 99.84 \\ \hline\hline
		B-PGM 0.05 & 2.50e+00 & 1.30e-02 & 2.06e-01 \\ \hline
		SY-RMI 0.05 & 2.24e+00 & 2.85e-02 & 8.52e-01 \\ \hline
		Best RMI & 1.00e+00 & 1.00e+00 & 1.00e+00 \\ \hline
		Best RS & 1.70e+00 & 2.40e+00 & 9.77e-01 \\ \hline
		\hline \multicolumn{4}{|c|}{L2}  \\ \hline \hline
        & Time & Space & Reduction Factor \\ \hline

		Best RMI & 3.32e+01 & 9.83e+01 & 99.98 \\ \hline\hline
		B-PGM 0.05 & 2.66e+00 & 1.02e-03 & 9.26e-01 \\ \hline
		SY-RMI 0.05 & 2.59e+00 & 1.63e-03 & 9.57e-01 \\ \hline
		Best RS & 1.60e+00 & 7.77e-02 & 9.97e-01 \\ \hline
		RMI $<$ 10 & 1.05e+00 & 2.50e-01 & 1.00e+00 \\ \hline
		\end{tabular}
		\qquad
        \begin{tabular}{|c|c|c|c|}
                \hline \multicolumn{4}{|c|}{L3}  \\ \hline \hline
                & Time & Space & Reduction Factor \\ \hline
        
		Best RMI & 5.16e+01 & 7.86e+02 & 100.00 \\ \hline\hline
		B-PGM 0.05 & 2.26e+00 & 3.76e-03 & 1.00e+00 \\ \hline
		SY-RMI 0.05 & 2.01e+00 & 3.83e-03 & 1.00e+00 \\ \hline
		Best RS & 1.74e+00 & 3.82e-02 & 1.00e+00 \\ \hline
		RMI $<$ 10 & 1.14e+00 & 5.00e-01 & 1.00e+00 \\ \hline
        \hline \multicolumn{4}{|c|}{L4}  \\ \hline \hline
        & Time & Space & Reduction Factor \\ \hline
        
		SY-RMI 2 & 1.61e-07 & 3.20e+04 & 100.00 \\ \hline\hline
		SY-RMI 0.05 & 1.39e+00 & 2.50e-02 & 1.00e+00 \\ \hline
		B-PGM 0.05 & 1.82e+00 & 2.53e-02 & 1.00e+00 \\ \hline
		Best RS & 1.30e+00 & 4.97e-01 & 1.00e+00 \\ \hline
		RMI $<$ 10 & 1.02e+00 & 7.86e-01 & 1.00e+00 \\ \hline
		\end{tabular}
		\end{center}
	\end{table}	
	
	\bibliographystyle{plain}
	\bibliography{references.bib}